\documentclass[aps,prb,twocolumn,superscriptaddress,floatfix]{revtex4-2}
\usepackage{algorithm}
\usepackage[noend]{algpseudocode}
\usepackage{amsmath}
\usepackage{amsfonts}
\usepackage{amssymb}
\usepackage{mathtools}
\usepackage{graphicx}
\usepackage{hyperref}
\usepackage{comment}
\usepackage{physics}
\usepackage{hyperref}
\usepackage[dvipsnames]{xcolor}
\usepackage{multirow}
\usepackage{xcolor}
\usepackage{amsthm}

\newtheorem{Pro}{Property}[section]
\usepackage{textcomp}

\begin{document}
\title{Quaternion-based machine learning on topological quantum systems}

\author{Min-Ruei Lin}
\email{m082030021@student.nsysu.edu.tw }
\affiliation{Department of Physics, National Sun Yat-sen University, Kaohsiung 80424,
Taiwan}

\author{Wan-Ju Li}
\affiliation{Department of Physics, National Sun Yat-sen University, Kaohsiung 80424,
Taiwan}

\author{Shin-Ming Huang}
\email{shinming@mail.nsysu.edu.tw}
\affiliation{Department of Physics, National Sun Yat-sen University, Kaohsiung 80424,
Taiwan}
\affiliation{Center of Crystal Research, National Sun Yat-sen University, Kaohsiung 80424, Taiwan}
\date{\today}

\begin{abstract}
Topological phase classifications have been intensively studied via machine-learning techniques where different forms of the training data are proposed in order to maximize the information extracted from the systems of interests. 
Due to the complexity in quantum physics, advanced mathematical architecture should be considered in designing machines.
In this work, we incorporate quaternion algebras into data analysis either in the frame of supervised and unsupervised learning to classify two-dimensional Chern insulators.
For the unsupervised-learning aspect, we apply the principal component analysis (PCA) on the quaternion-transformed eigenstates to distinguish topological phases.
For the supervised-learning aspect, we construct our machine by adding one quaternion convolutional layer on top of a conventional convolutional neural network.
The machine takes quaternion-transformed configurations as inputs and successfully classify all distinct topological phases, even for those states that have different distributuions from those states seen by the machine during the training process.
Our work demonstrates the power of quaternion algebras on extracting crucial features from the targeted data and the advantages of quaternion-based neural networks than conventional ones in the tasks of topological phase classifications. 
\end{abstract}
\maketitle

\section{Introduction}
The phase classification using machine-learning (ML) based techniques has been attracting intense attentions since the pioneering work in 2017 \cite{carrasquilla2017}.
In addition to the classical phase detections \cite{bedolla2021,mehta2019} where each phase is well defined by the corresponding order parameters, detecting topological phase transitions \cite{bedolla2021} is interesting and challenging \cite{beach2018} due to the lack of local order parameters.
Recently, the phase detections and classifications have been performed via different ML techniques for classifying various topological invariants \cite{yoshioka2018,carvalho2018,balabanov2020,balabanov2021,greplova2020,greplova2020,ho2021,zhang2021,narayan2021,yu2021,
zhang2017,cheng2018,sun2018,zhang2020,carvalho2018,kerr2021,kaming2021,ho2021,rem2019,che2020,chung2021,ming2019quantum,zhang2018,sun2018,
holanda2020,che2020,kerr2021,tsai2021,kerr2021,zhang2020,zhang2017a,mano2019,su2019,lian2019,carvalho2018,ho2021,beach2018,laskowska2018,zhang2019,
RN2019,tsai2020,scheurer2020,tsai2021,caio2019}, including the Chern number \cite{zhang2017,cheng2018,sun2018,zhang2020,carvalho2018,kerr2021,kaming2021,ho2021,rem2019,che2020,chung2021,ming2019quantum}, winding number \cite{zhang2018,sun2018,holanda2020,che2020,kerr2021,tsai2021,kerr2021}, $\mathbb{Z}_2$ index \cite{zhang2020,zhang2017a,mano2019,su2019,lian2019,carvalho2018,ho2021,beach2018,laskowska2018,zhang2019,
RN2019,tsai2020,scheurer2020,tsai2021,caio2019}, to name a few.
In addition to the applied ML architectures, the forms of the inputs for training the machine also play a crucial role in determining the resulting performance of the topological phase detections \cite{beach2018}.

For the topological systems with the Chern numbers or the winding numbers as the topological invariants, various types of inputs are used to perform the phase classifications.
For instance, the quantum loop topography (QLT) is introduced to construct multi-dimensional images from raw Hamiltonians or wave functions as inputs \cite{zhang2017,zhang2020}. The Bloch Hamiltonians are arranged into an arrays to feed the neural networks \cite{zhang2018,sun2018}. In addition, the real-space particle densities and local density of states \cite{cheng2018} and the local projections of the density matrix \cite{carvalho2018} are also used as inputs.
From cold-atom experiments, momentum-space density images were generated as inputs for classifications \cite{rem2019}.
The time-of-flight images \cite{ho2021,kaming2021}, spatial correlation function \cite{ho2021}, density–density correlation function \cite{ho2021} and 
the density profiles formed in quantum walks were also proposed as appropriate inputs \cite{ming2019quantum}.
Furthermore, the spin configurations \cite{kerr2021} and the Bloch Hamiltonians over the Brillouin zone (BZ) have also been treated as inputs for the neural networks \cite{che2020,kerr2021}.
For these forms of inputs mentioned above, various ML techniques with distinct real-valued neural networks have been applied to discriminate different topological phases.

As the development of artificial neural networks becomes mature, a raise of representation capability of machines is anticipated by generalizing real-valued neural networks to complex-valued ones~\cite{trabelsi2017,gaudet2018deep}.
Specifically, a quaternion number, containing one real part and three imaginary parts, and the corresponding quaternion-based neural networks \cite{GR2020,isokawa2009,parcollet2020,matsui2004} are expected to enhance the performance on processing of data with more degrees of freedom than the conventional real-number and complex-number systems.
There have been various proposals about quaternion-based neural networks in ML techniques and applications in computer science, such as the quaternion convolutional neural network (qCNN) \cite{zhu2019,hongo2020,gaudet2018deep}, quaternion recurrent neural network \cite{parcollet2018quaternion}, quaternion generative adversarial networks \cite{grassucci2021}, quaternion-valued variational autoencoder \cite{grassucci2020}, quaternion graph neural networks \cite{nguyen2020}, quaternion capsule networks \cite{ozcan2020} and quaternion neural networks for the speech recognitions \cite{parcollet2018}.
However, the ML-related applications of the quaternion-based neural networks on solving problems in physics are still limited, especially in the topological phase detections, even though the quaternion-related concepts have been applied in some fields in physics~\cite{girard1984,girard2007,girard2018}.
\begin{figure}[th!]
    \centering
    \includegraphics[width=0.46\textwidth]{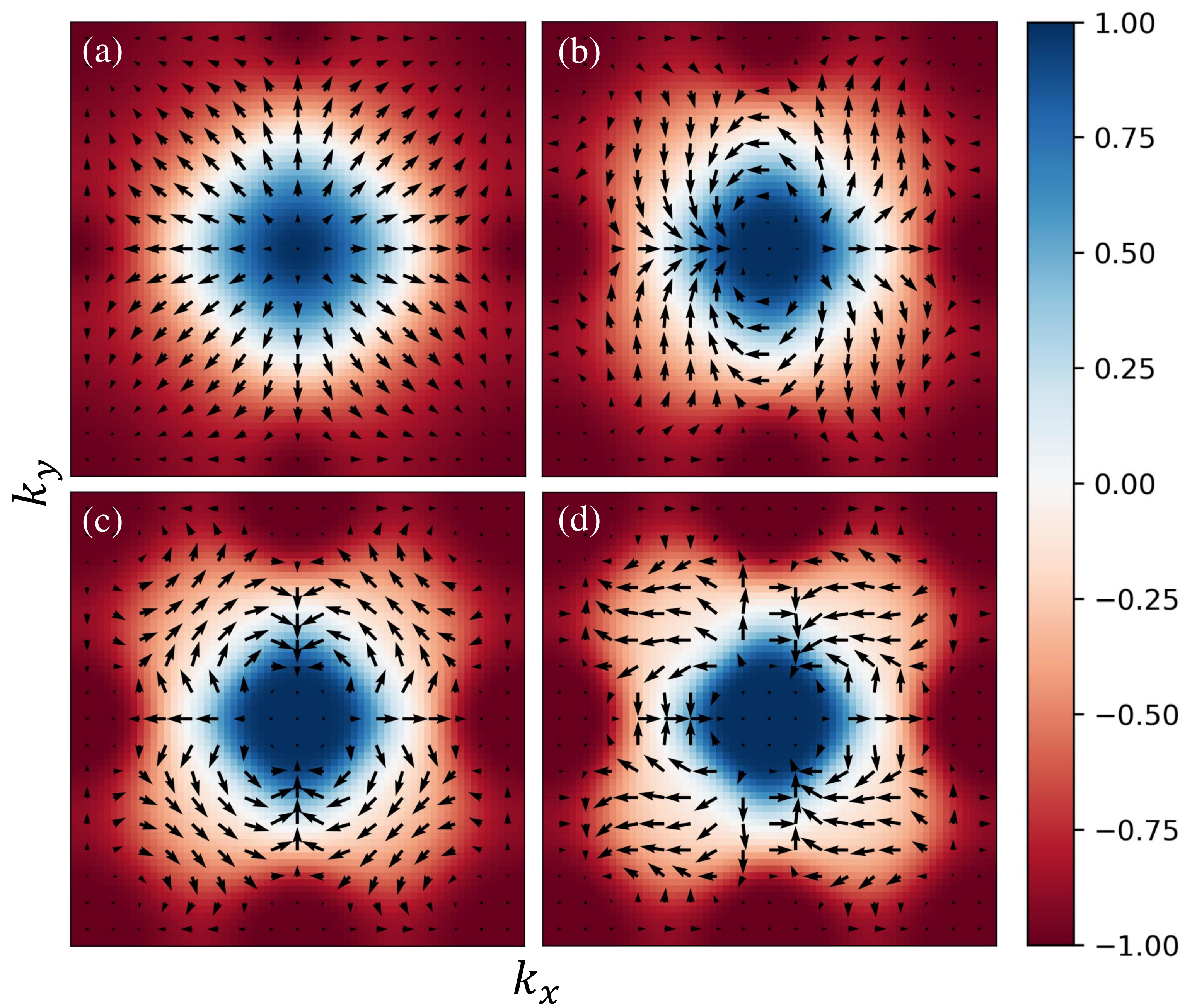}
    \caption{Examples of spin textures in the Brillouin zone, $k_x,k_y\in (-\pi,\pi]$, with the Chern number \(C = 1\)  (a), \(C=2\) (b), \(C = 3\) (c) and \(C=4\) (d).}
    \label{fig:spin_conf}
\end{figure}

In this work, we perform the Chern-insulator classifications from both supervised- and unsupervised-learning aspects based on the inputs transformed via the quaternion algebra.
For the unsupervised learning, we encode the quaternion-transformed eigenstates of Chern insulators via a convolution function as inputs and study them using the principal component analysis (PCA).
We found that using only the first two principal elements is not enough to fully classify the Chern insulators, consistent with Ming's work \cite{ming2019quantum}.
Further studies show that the performance can be improved by including more principal components.
For the supervised learning, we construct a quaternion-based neural network in which the first layer is a quaternion convolutional layer.
We then show that this quaternion-based machine has better performance than a conventional CNN machine. Our machine is good not only for testing datasets but also for identifying data points that have different distributions from those seen by our machine in the training processes.
The good performance can be attributed to the similarities between the formula of the Berry curvatures and our quaternion-based setup.
Therefore, our work demonstrates the power of the quaternion algebra on extracting relevant information from data, paving the way to applications of quaternion-based ML techniques in topological phase classifications.

The outline of the remaining part of this work is as follows.
In Sec. II, we introduce the model Hamiltonian, generating the data for our classification tasks, and the quaternion convolution layer used in this work.
PCA analysis of the quaternion-transformed eigenstates is discussed in Sec. III.
The data preparations, the network structures and the performance of the quaternion-based supervised learning task are given in Sec. IV.
Discussions and Conclusions are presented in Sec. V and Sec. VI, respectively. We have three appendixes. Appendix A shows the details of data preparation. Appendix B provides a brief introduction to the quaternion algebra. Some properties of functions in Sec. III are included in Appendix C. 

\section{Model and quaternion convolutional layer}
\subsection{Model}
A generic two-band Bloch Hamiltonian with the aid of the identity matrix $\sigma_0$ and Pauli matrices \(\pmb{\sigma}=(\sigma_1, \sigma_2,\sigma_3)\) is written as
\begin{equation}\label{eq:2band}
    \mathcal{H}(\vec{k}) = h_0(\vec{k}) \sigma_0 + \mathbf{h}(\vec{k})\cdot\pmb{\sigma},
\end{equation}
where $\vec{k}=(k_x,k_y)$ is the crystal momentum in the 2D BZ ($\forall {k_x,k_y\in(-\pi,\pi]}$). $h_0(\vec{k})$ can change energy of the system but has nothing to do with topology, so it will be ignored in the remaining part of this paper. The vector \(\mathbf{h} = (h_1,h_2,h_3)\) acts as an $k$-dependent external magnetic field to the spin $\vec{\sigma}$, so that the eigenstate of the upper (lower) band at each $\vec{k}$ will be the spin pointing antiparallel (parallel) to $\mathbf{h}(\vec{k})$. It will be reasonable that the unit vector $\mathbf{n} = \mathbf{h}/|\mathbf{h}| \in S^2$ embeds the topology in this system. Indeed, the topological invariant is the Chern number C,
\begin{equation}\label{chern}
    C = \frac{1}{4\pi}\int_{\text{BZ}} \mathbf{n}\cdot(\partial_{k_x}\mathbf{n}\times\partial_{k_y}\mathbf{n})d\vec{k},
\end{equation}
where the integrand is the Berry curvature and the integration is over the first BZ. For brevity, sometimes we will omit the argument $\vec{k}$ in functions.
The Chern number is analogous to the skyrmion number in real space \cite{nagaosa2013}. The integral is the total solid angle $\mathbf{n}(\vec{k})$ subtended in the BZ, so the Chern number counts how many times $\mathbf{n}(\vec{k})$ wraps a sphere.
\begin{figure}[b]
    \centering
    \includegraphics[width=0.45\textwidth]{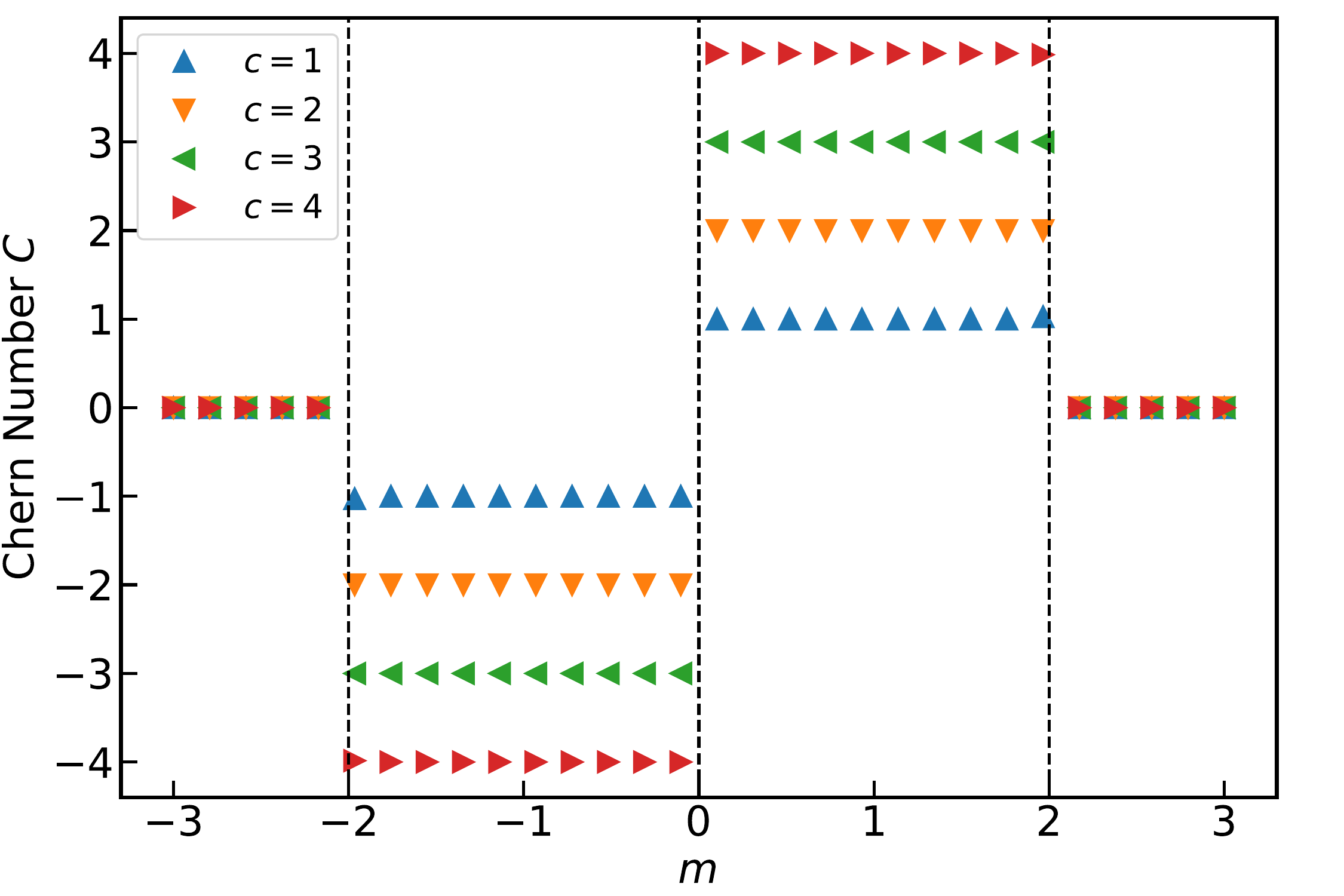}
    \caption{The Chern number with various \(m\) and \(c\).}
    \label{fig:HighChern}
\end{figure}

We construct the normalized spin configurations $\mathbf{n}(\vec{k})$ based on the following models. For topological systems, we choose the Hamiltonian with \(\mathbf{h}=\mathbf{h}^{(c)}\), where
\begin{equation}\label{eq:HighChern}
\mathbf{h}^{(c)}(\vec{k},m)=
    \begin{pmatrix}
        \ \mathrm{Re}\big[(\sin{k_x}-i\sin{k_y})^c\big]\\
        -\mathrm{Im}\big[(\sin{k_x}-i\sin{k_y})^c\big]\\
        \cos{k_x} + \cos{k_y} + m
    \end{pmatrix}
\end{equation}
with positive integer \(c\) and real parameter \(m\) to control the Chern number. \textit{c} is the vorticity for the number of times the inplane component ($n_x$ and $n_y$) swirls around the origin. The sign of the \textit{c} indicates a counter-clockwise or clockwise swirl. For a nontrivial topology, $n_z$ has to change sign somewhere in the BZ for $\mathbf{n}(\vec{k})$ to wrap a complete sphere. Therefore, $|m|<2$ is required. Some examples of spin texture $\mathbf{n}(\vec{k})$ based on Eq. (\ref{eq:HighChern}) are shown in Fig.~\ref{fig:spin_conf}. 
For \(c = 1\), the model is the Qi-Wu-Zhang (QWZ) model \cite{PhysRevB.74.085308}. For a given \(c\), the Chern number $C$ can be either \(0,~c,\text{ or} -c\) depending on the value of \(m\): 
\begin{equation}
    C =
    \begin{cases} 
    \mathrm{sgn}(m)c, & 0<|m|<2, \\
    0, & |m|>2.
    \end{cases}
\end{equation}
The topological phase diagram is shown in Fig.~\ref{fig:HighChern}. $C=0$ denotes a topologically trivial phase and $C \neq 0$ a nontrivial phase.

In this work, the unsupervised learning involves seven topological phases ($C=0,\pm 1,\pm 2, \pm 3$) in Sec. \ref{sec:PCA}, and the supervised learning involves nine topological phases ($C=0,\pm 1,\pm 2, \pm 3, \pm 4$) in Sec. \ref{sec:qcnn}.

\subsection{Quaternion convolutional layer}
A quaternion number has four components, the first of which stands for the real part and the other three of which stand for the imaginary parts. Given two quaternions $q_1=(r_1,a_1,b_1, c_1)$ and $q_2=(r_2,a_2,b_2,c_2)$, their product $Q=q_1q_2=(R,A,B,C)$ is given by
\begin{equation}\label{eq:q_1q_2}
    \begin{pmatrix}
        R\\
        A\\
        B\\
        C
    \end{pmatrix} =
    \begin{pmatrix}
        r_1 r_2 - a_1 a_2 - b_1 b_2 - c_1 c_2\\
        a_1 r_2 + r_1 a_2 - c_1 b_2 + b_1 c_2\\
        b_1 r_2 + c_1 a_2 + r_1 b_2 - a_1 c_2\\
        c_1 r_2 - b_1 a_2 + a_1 b_2 + r_1 c_2
    \end{pmatrix},
\end{equation}
which can be written as the matrix product form
\begin{equation}\label{eq:matrix_q}
\begin{pmatrix}
        R\\
        A\\
        B\\
        C
    \end{pmatrix} =
    \begin{pmatrix*}[r]
        r_1 & -a_1 & -b_1 & -c_1 \\
        a_1 & r_1 & -c_1 & b_1 \\
        b_1 & c_1 & r_1 & -a_1 \\
        c_1 & -b_1 & a_1 & r_1
    \end{pmatrix*}
    \begin{pmatrix}
        r_2\\
        a_2\\
        b_2\\
        c_2
    \end{pmatrix}.
\end{equation}
To implement a quaternion convolutional (q-Conv) layer in numerical programming, we will regard the two quaternions as a $4\times4$ matrix and a $4\times1$ column matrix, respectively:
\begin{equation} \label{eq:rep2}
q_1 \doteq 
    \begin{pmatrix*}[r]
        r_1 & -a_1 &-b_1 & -c_1 \\
        a_1 & r_1 & -c_1 & b_1 \\
        b_1 & c_1 & r_1 & -a_1 \\
        c_1 & -b_1 & a_1 & r_1
    \end{pmatrix*}\quad \mathrm{and}\quad
    q_2 \doteq 
    \begin{pmatrix*}[r]
        r_2\\
        a_2\\
        b_2\\
        c_2
    \end{pmatrix*}.
\end{equation}
More details of quaternion algebra are described in Appendix~\ref{app:qcnn}.

A conventional CNN contains a real-valued convolutional layer to execute the convolution of the input and the kernel. 
Let the input $F$ have the shape: $H_i\times W_i\times C_i$ (Height \(\times\) Width \(\times\) Channel) and the shape of the kernel $K$ be $H_{k} \times W_{k} \times C_i \times C_f$.
The convolution will produce an output \(O\), $O=F\ast K$, whose elements are
\begin{equation}\label{eq:C-convo}
    O_{i',j',t'} =\sum_{i}^{H_k} \sum_{j}^{W_k} \sum_{t}^{C_i} F_{i'+i-1,j'+j-1,t}\cdot K_{i,j,t,t'}.
\end{equation}
Here the stride is assumed to be \(1\) both in the width and the height directions.
The indices $i$ and $j$ are spatial indicators, $t$ is the index of channel in the input feature map and $t'$ is the kernel index.
The shape of the output will be $(H_{i}-H_{k}) \times (W_{i}-W_{k}) \times C_f$.

Assume that the input has four components. To uncover the entanglement among components through CNN, we will utilize the quaternion product. Now, we introduce another dimension–depth–which is four, as a quaternion number of four components. Both of the input $F$ and the kernel $K$ have depth of four as two quaternion numbers. The product of $F$ and $K$ will have depth of four as a quaternion in Eq.~(\ref{eq:q_1q_2}). Referring to Eq.~(\ref{eq:rep2}) where we show a matrix representation to implement quaternion algebra and thinking of $F$ as $q_1$ and $K$ as $q_2$ in Eq.~(\ref{eq:rep2}), we transform the depth-four input $F$ into a $4\times 4$ matrix, $F^{(s,l)}$, and keep the kernel $K$ still of depth 4, $K^{(l)}$, where $l, s=1,…,4$. The product of $F$ and $K$, say $O$, will have depth four as shown in Eq.~(\ref{eq:Q-convo}). Further considering the shapes of $F$ and $K$, the convolution is given by
\begin{equation}\label{eq:Q-convo}
    O_{i',j',t'}^{(s)} = \sum_{l}^{4}\sum_{i,j,t} F^{(s,l)}_{i'+i-1,j'+j-1,t}\cdot K^{(l)}_{i,j,t,t'},
\end{equation}
where the summations over $i$,$j$,$k$ are equivalent to those in Eq.~(\ref{eq:C-convo}) and the summation over $l$ is for the quaternion product.
\begin{figure}[b!]
    \centering
    \includegraphics[width=0.47\textwidth]{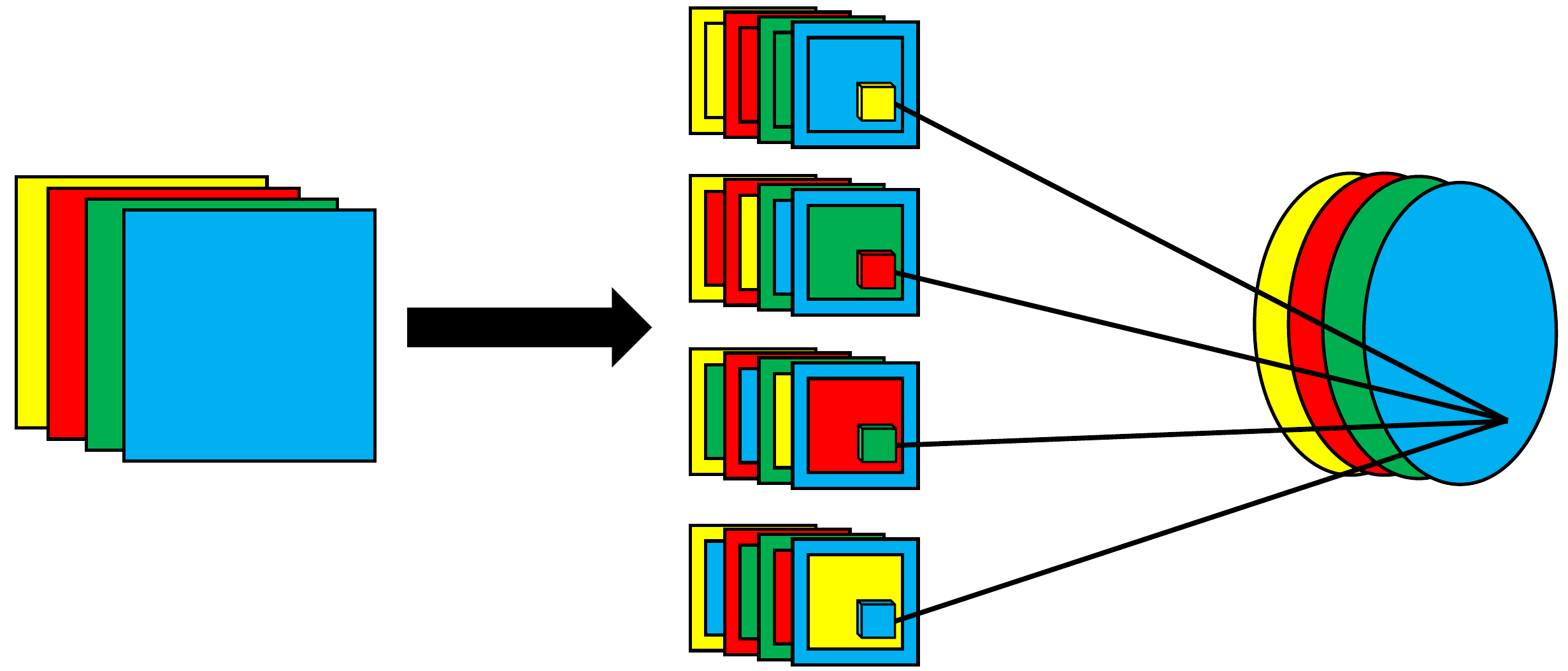}
    \caption{Illustration of a quaternion convolutional layer. On the left, we start with the input $q_1$ having four quaternion components ((yellow, red, green, blue) stands for ($r_1$, $a_1$, $b_1$, $c_1$)). In the middle, $q_1$ is permuted to construct $\{F^{(\cdot, l)}\}_{l=1}^{4}$ on which the convolution with four kernels $\{K^{(l)}\}_{l=1}^{4}$ is performed. A summation is taken for each depth to obtain the output feature map $O$ on the right.} 
    \label{fig:Q-filter}
\end{figure}

More specifically, we consider an input data as $q_1$ (four color squares on the left of Fig. \ref{fig:Q-filter}) and four kernels encoded in $q_2$, given in the following
\begin{equation}
\left\{\begin{matrix}
q_1 \doteq & (r_1~a_1~b_1~c_1)^T &\\
q_2 \doteq & (r_2~a_2~b_2~c_2)^T & =: K^{(\cdot)}.
\end{matrix}\right.
\end{equation}
The output feature maps $O\doteq (R~A~B~C)^T$ is then calculated based on Eq.~(\ref{eq:q_1q_2}).
As the first step, we permute the order of $q_1$ to obtain
\begin{eqnarray}
F^{(\cdot,1)}=:
\begin{pmatrix}
        r_1\\
        a_1\\
        b_1\\
        c_1
    \end{pmatrix},
F^{(\cdot,2)}=:
\begin{pmatrix*}[r]
        -a_1\\
        r_1\\
        c_1\\
        -b_1
    \end{pmatrix*},\\\nonumber
F^{(\cdot,3)}=:
    \begin{pmatrix*}[r]
        -b_1\\
        -c_1\\
        r_1\\
        a_1
    \end{pmatrix*},
F^{(\cdot,4)}=:
    \begin{pmatrix*}[r]
        -c_1\\
        b_1\\
        -a_1\\
        r_1
    \end{pmatrix*}
\end{eqnarray}
(see the four sets of sqaures in the middle of Fig. \ref{fig:Q-filter}). We then convolute those four quaternions ($F^{(\cdot,l)}$ with $l =1,2,3$ and 4) with four kernels ($K^{(l)}$ with $l = 1,2,3$ and 4) in the following way:
$$
\left\{
\begin{matrix}
    F^{(\cdot,1)}K^{(1)} &\doteq& 
        \begin{pmatrix}
            r_1r_2&a_1r_2&b_1r_2&c_1r_2
        \end{pmatrix}^T\\
    
    F^{(\cdot,2)}K^{(2)} &\doteq&
        \begin{pmatrix}
            -a_1a_2&r_1a_2&c_1a_2&-b_1a_2
        \end{pmatrix}^T\\
    
    F^{(\cdot,3)}K^{(3)} &\doteq&
        \begin{pmatrix}
            -b_1b_2&-c_1b_2&r_1b_2&a_1b_2
        \end{pmatrix}^T\\
    
    F^{(\cdot,4)}K^{(4)} &\doteq& 
        \begin{pmatrix}
            -c_1c_2&b_1c_2&-a_1c_2&r_1c_2
        \end{pmatrix}^T
\end{matrix}
\right.
$$
as shown in the middle of Fig. \ref{fig:Q-filter}. Finally, we sum over the above four quaternions to get the output feature maps $O$, as shown on the right of Fig. \ref{fig:Q-filter}.
$$
O
:=
\begin{pmatrix}
    R\\
    A\\
    B\\
    C
\end{pmatrix}
=
\begin{pmatrix}
    r_1 r_2 - a_1 a_2 - b_1 b_2 - c_1 c_2\\
    a_1 r_2 + r_1 a_2 - c_1 b_2 + b_1 c_2\\
    b_1 r_2 + c_1 a_2 + r_1 b_2 - a_1 c_2\\
    c_1 r_2 - b_1 a_2 + a_1 b_2 + r_1 c_2    
\end{pmatrix}.
$$

\section{principal component analysis}\label{sec:PCA}
Principal component analysis (PCA) is a linear manifold learning that is to find the relevant basis set among data \cite{jolliffe2016principal,ma2012manifold}.

We prepare eigenstates $\ket{u_\pm}$ of Eq.~(\ref{eq:2band}), where $+\ (-)$ stands for the upper (lower) band.
For a topologically nontrivial state, the phase cannot be continuous over the whole BZ. Therefore, we can divide the whole BZ into two parts, in each part of them the topological wave function has continuously well-defined phase. We then pick up a gauge by choosing two regions according to the sign of $h_3$ in Eq. (\ref{eq:HighChern}):
\begin{align}
    \begin{split}
         \ket{u_+} & \doteq \frac{1}{\sqrt{2h_+(h_{+}+h_3)}}\mqty(h_{+}+h_3 \\ h_1+i h_2) \\
         \ket{u_-} & \doteq \frac{1}{\sqrt{2h_-(h_{-}+h_3)}} \mqty(-h_1+ih_2 \\ h_{-}+h_3)
         \end{split}
         \quad ,~ h_3\geq 0,  \label{eq:gauge1}
\end{align}
and
\begin{align}
 \begin{split}
 \ket{u_+} & \doteq \frac{1}{\sqrt{2h_{+}(h_{+}-h_3)}}\mqty(h_1-ih_2 \\  h_{+}-h_3) \\
 \ket{u_-} & \doteq \frac{1}{\sqrt{2h_{-}(h_{-}-h_3)}} \mqty(h_{-}-h_3 \\ -h_1-ih_2)
 \end{split}
 \quad ,~ h_3 < 0, \label{eq:gauge2}
\end{align}
where $h_{\pm} =\pm \sqrt{h_1^2+h_2^2+h_3^2}$. In this choice of gauge, the first (second) component of $\ket{u_{+}}$ $(\ket{u_-})$ is real-valued when $h_3\geq0$, and the second (first) component of $\ket{u_{+}}$ $(\ket{u_-})$ is real-valued when $h_3<0$.

By translating $\ket{u_\pm} \doteq (\alpha_\pm,  \beta_\pm)^T$ with $\alpha_\pm, \beta_\pm\in\mathbb{C}$, into a quaternion number of four components, we have
\begin{equation}\label{eq:transf}
    q_\pm := \mathrm{Re}(\alpha_\pm) + \mathrm{Im}(\alpha_\pm)\hat{\mathbf{i}} + \mathrm{Re}(\beta_\pm)\hat{\mathbf{j}} + \mathrm{Im}(\beta_\pm)\hat{\mathbf{k}}.
\end{equation}
To see the correlation of states over $\vec{k}$, we define the quantity $F$ to be the quaternion-based convolutions:
\begin{equation}\label{eq:Ne}
\begin{split}
    F(\vec{p}) :=  (q^*_{+}&\circledast q_{+})[\vec{p}] - (q^*_{-}\circledast q_{-})[\vec{p}]
    \quad\mathrm{with}\\
    (q_{\pm}^*&\circledast q_{\pm})[\vec{p}] :=  \sum_{\vec{k}\in \mathrm{BZ}} q_{\pm}^*(\vec{k})q_{\pm}(\vec{p} - \vec{k}),
\end{split}
\end{equation}
where $q^*$ is the conjugate of $q$. It can be proved that $F$ is real-valued. Therefore, $F(\vec{p})$ of all $\vec{p}$ in the BZ based on a given Hamiltonian can be analysed by using PCA.
\begin{figure}[b!]
    \centering
    \includegraphics[width=0.45\textwidth]{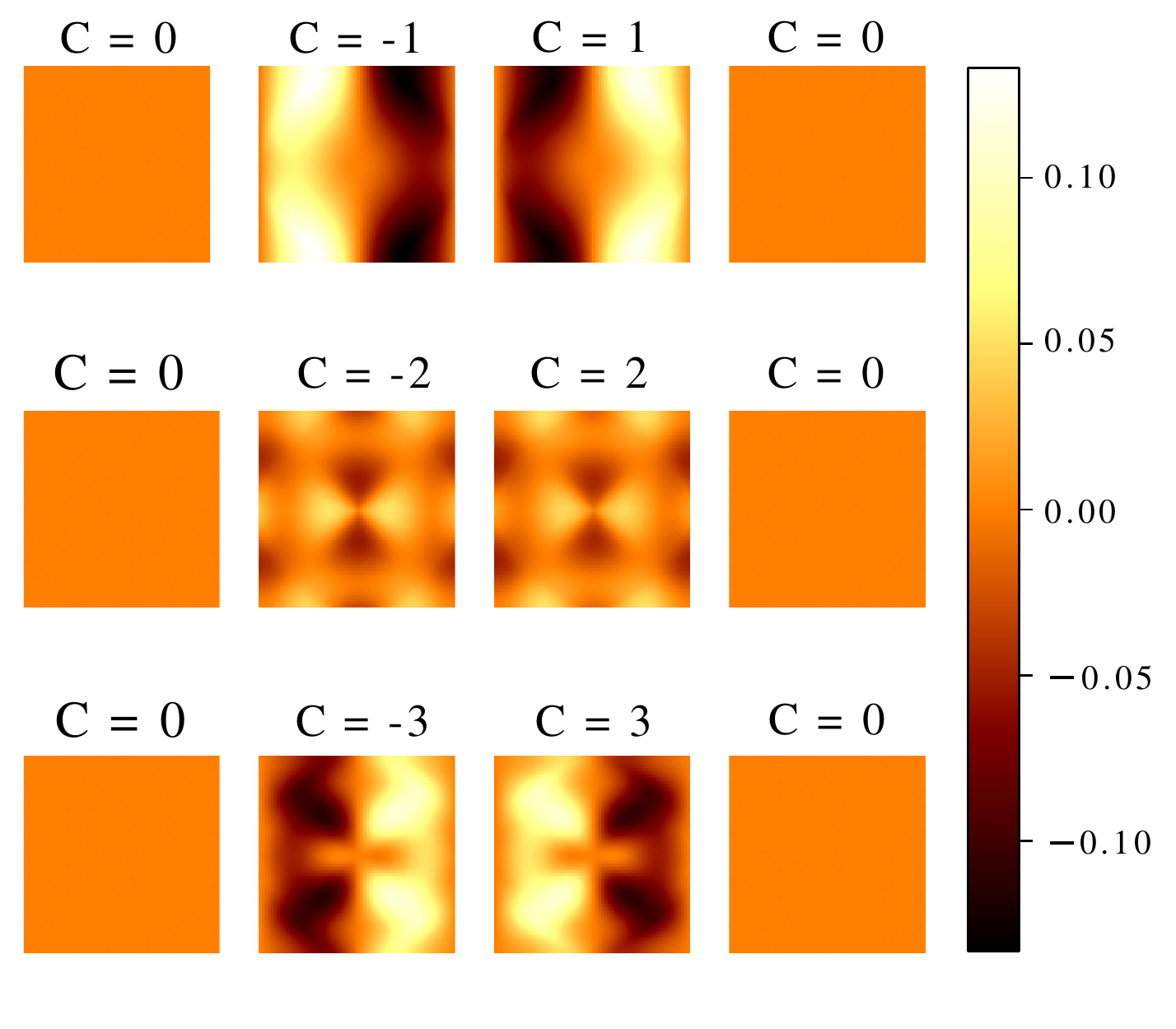}
    \caption{The maps of the function \(F\) without noise in the BZ. Three rows are for \(c = 1, 2\ \mathrm{and}\ 3\) in Eq.~\ref{eq:HighChern} from top to bottom; four columns from left to right are for $m = -3,-1, 1$ and 3. The corresponding Chern number $C$ is tagged with each panel.}
    \label{fig:nematic}
\end{figure}

We collected various $F$ of all $\vec{k} \in $BZ within seven topological phases as the dataset for PCA.
For each topological phase, 30 $F$'s were prepared, so the total amount of data was 210.
The data for six non-trivial phases were generated based on Eq. (\ref{eq:HighChern}) with $m=\pm 1$ (the sign of $m$ determines the sign of $C$).
For the trivial phase, we prepared five data points from each of six combinations of $\{c,m\}$, where $c\in\{1,2,3\}$ and $m\in\{3,-3\}$, and then there are totally 30 data.
To augment the number of data, we add Gaussian noises $\delta\mathbf{h}$ at every $\vec{k}$ of the model [Eq. (\ref{eq:HighChern})] such that $\mathbf{h}\to\mathbf{h}+\delta\mathbf{h}$ without closing the band gap.

In Fig.~\ref{fig:nematic}, we present various noiseless $F$ generated from Eq. (\ref{eq:HighChern}) with different $c$ and $m$.
\begin{figure}[t!]
    \centering
    \includegraphics[width=0.49\textwidth]{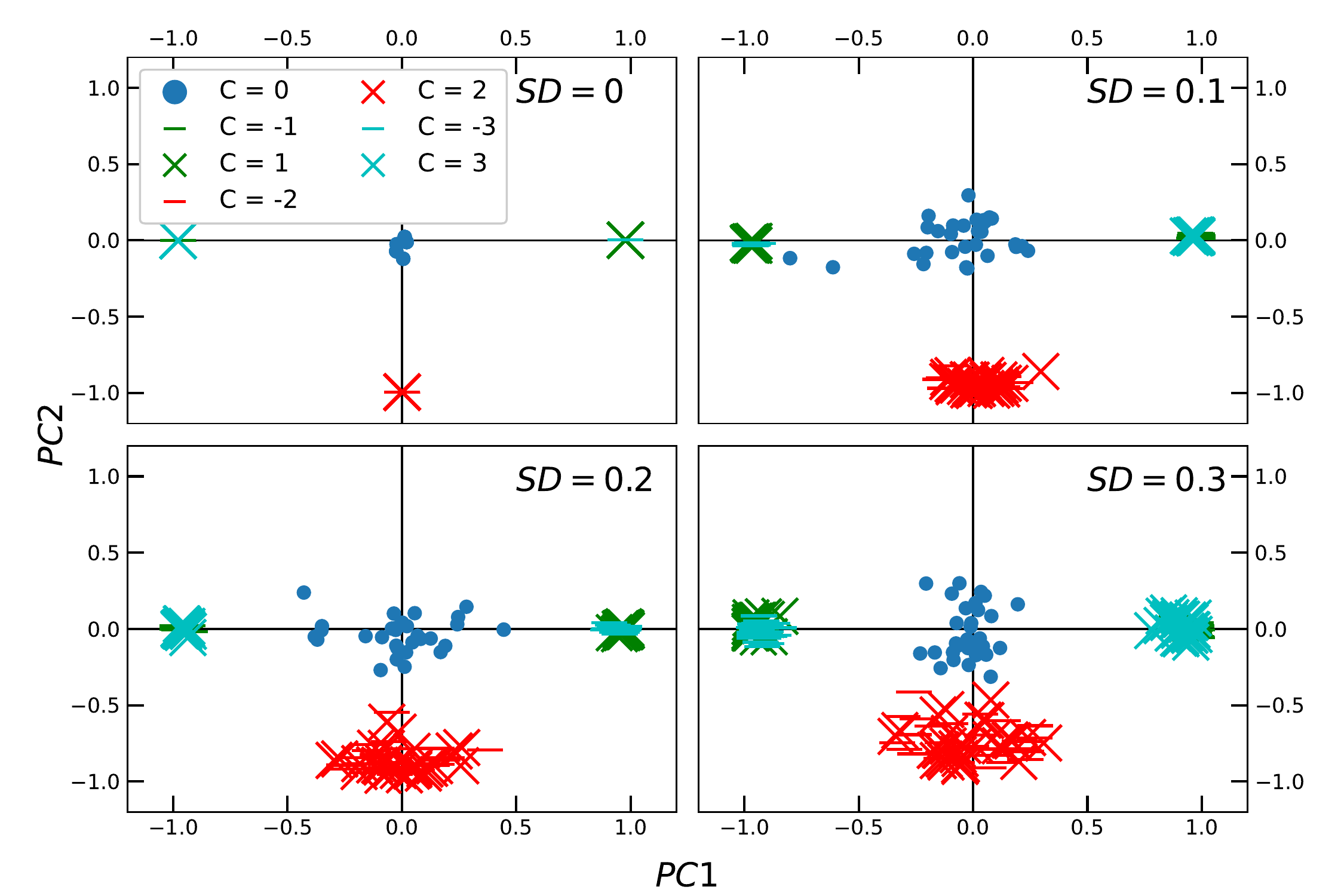}
    \caption{PCA of seven topological phases with various noise. The symbols with corresponding Chren numbers are marked in the legend.}
    \label{fig:2DPCA}
\end{figure}
It is notable that $F$ for $C = 0$ are featureless, $F$ for $C = \pm 1$ have a dipole moment, and $F$ for $C = \pm 2$ have a quadruple moment, and $F$ for $C = \pm 3$ seemingly have a primary dipole and a secondary quadruple moment. 
The remarkable features imply that the convolution function $F$ is a good choice for topological classifications.

We examine data with the standard deviation (SD) equal to 0, 0.1, 0.2 and 0.3 respectively, and show the first two PCs of 210 pieces of data for each SD in Fig.~\ref{fig:2DPCA}. In Fig.~\ref{fig:2DPCA}, data are clustered into four groups and their variances increase with SD. It is successful to separate different topological phases into different clusters via PCA. However, some clusters contain two topological phases of Chern numbers: $\{+1,-3\}$, $\{-1,+3\}$, and $\{+2,-2\}$. This $C$ modulo 4 resemblance has also be observed in a previous study~\cite{ming2019quantum}.

\begin{figure}[b!]
    \centering
    \includegraphics[width=0.5\textwidth]{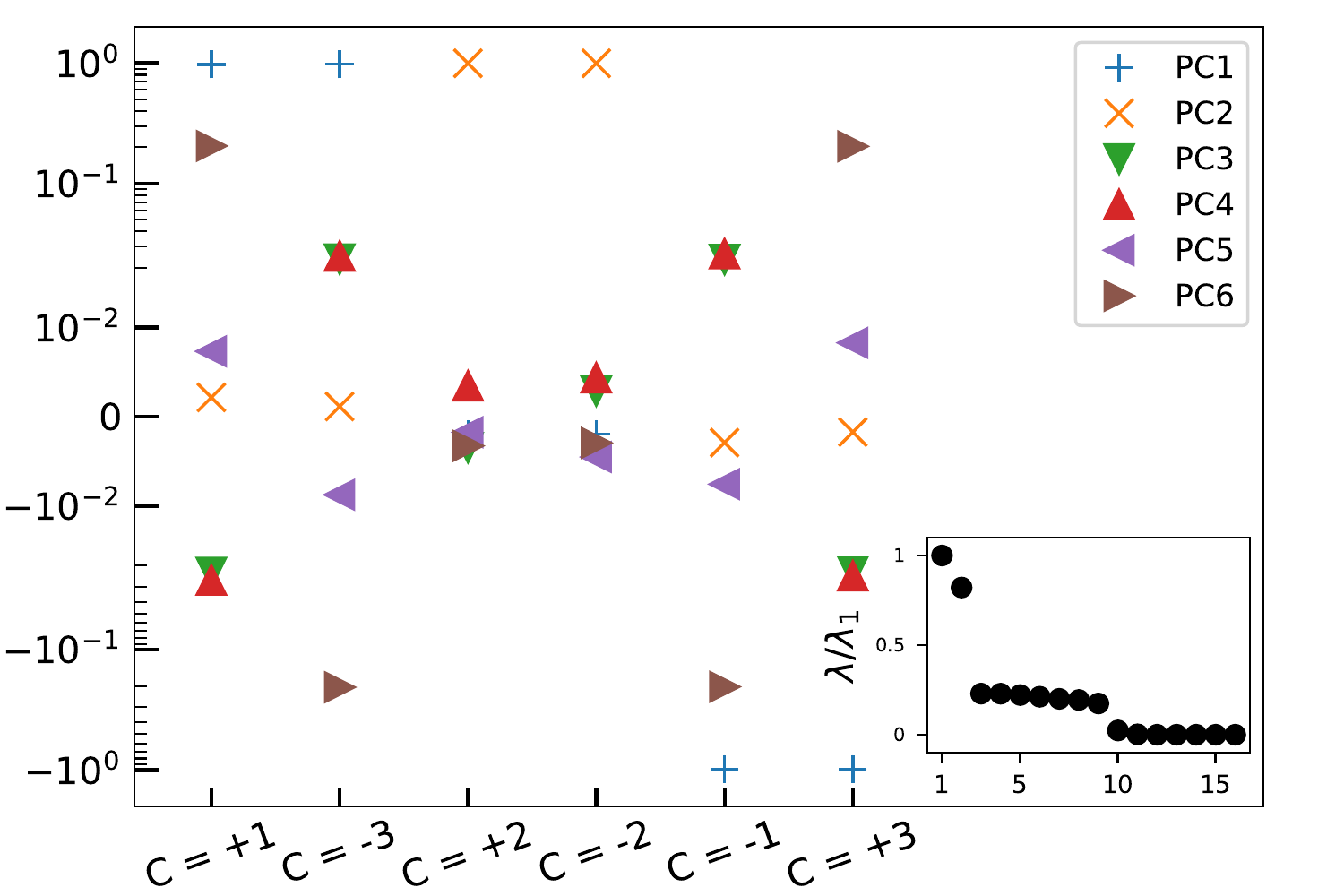}
    \caption{Magnitude of projection (logarithmic scale) from non-trivial data onto first six principal components. Inset: The first 16 principal values of PCA. (normalized by maximal $\lambda_1$)}
    \label{fig:ninePC}
\end{figure}
We find that including more PCs helps separate different classes in each cluster. Figure \ref{fig:ninePC} shows the first six PCs of data in topologically non-trivial phases, where PC$x$ denotes the $x$-th PC component. One can find that PC1 and PC2 in each pair of $\{+1,-3\}$, $\{-1,+3\}$, and $\{+2,-2\}$ are nearly identical, as also shown in Fig.~\ref{fig:2DPCA}. By incorporating more PCs up to PC6, all topological classes are completely classified. Via the proposed convolution, topological states can be successfully classified by using PCA, a linear classification machine.

Compared to the eigenstates, the spin configurations $\mathbf{n(}\vec{k})$ are gauge-invariant. Therefore, it is desired to classify the topology of the spin configurations via PCA. Unfortunately, the performance was not good, which will be discussed later. In order to directly classify the spin configurations, in the following, we train a qCNN machine via the supervised learning algorithm to discriminate spin configurations with different topological phases.

\begin{figure*}[t!]
    \centering
    \includegraphics[width=0.945\textwidth]{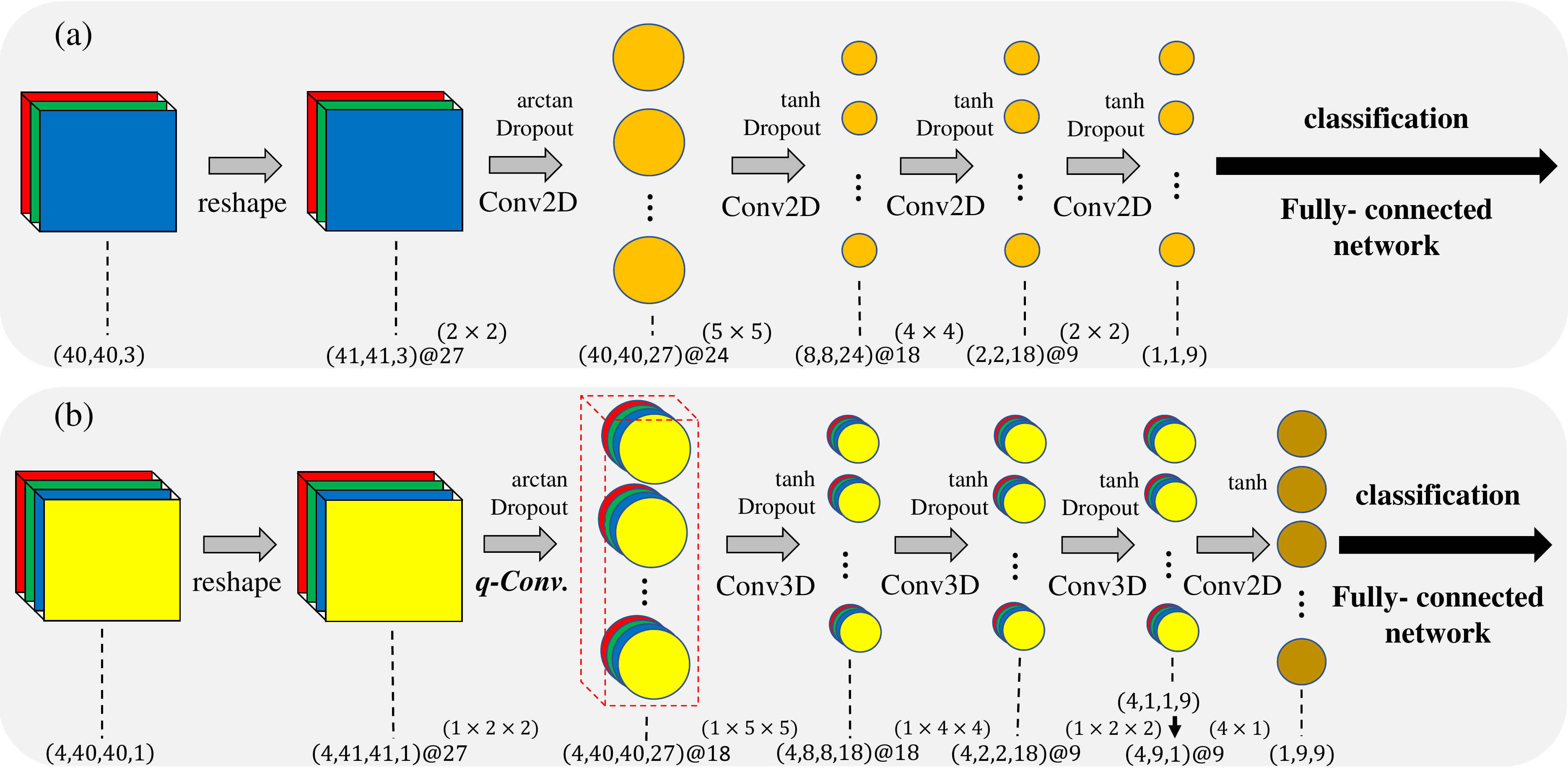}
    \caption{Framework of (a) the CNN and (b) the qCNN classifier.
    In CNN, the shapes of data are labelled as (Width, Height, Channel) and the spin components are arranged in different channels in the first layer, while in qCNN the shapes of data are labelled as (Depth, Width, Height, Channel) but spin components are put along depth. The number appended after the symbol ``@'' stands for the number of kernels, which size is present overhead.
    The number of tuneable network parameters of CNN (qCNN) is 24,252 (19,350).}
    \label{fig:QCNN}
\end{figure*}
\section{Supervised learning of CNN and the qCNN}\label{sec:qcnn}
\subsection{Datasets}
The input data are normalized spin configurations $\mathbf{n}$, laying on a $40\times 40$ square lattice with periodic boundary conditions, and their corresponding topological phases are labels with one-hot encoding.
We prepared four datasets: training, validation, testing and prediction dataset (more details are described in Appendix \ref{app:dataset}).

The first three datasets are well known in conventional deep learning procedure~\cite{chollet2021deep}. The data in the training, validation and testing datasets will be constructed by the same models so that they have the same data distributions even though they are all different data points. Therefore, we denote these three datasets as in-distribution datasets. The data in the prediction dataset, however, are constructed by similar but different models from those for the in-distribution datasets. Therefore, the data in the prediction dataset are not only unseen by the machine during the training process, but also of different distributions. We denote the prediction dataset as a out-of-distribution dataset, which is used to understand whether our machine can also classify spin configurations constructed by other similar but different topological models.

The data pool containing training and validation datasets is constructed as follows.
Based on the Eq.~(\ref{eq:HighChern}), we firstly prepared 5760 data points of $\mathbf{n}$ of nine topological phases with Chern number ranging from -4 to 4 and each phase contains 640 data points.
Besides of 5760 spin configurations, the dataset contains 360 two-dimensional spin vortices.
A spin-vortex has an in-plane spin texture that winds around a center, which is generated by setting one of three components in Eq.~(\ref{eq:HighChern}) to be zero.
By including spin vortices, the machine can tell the difference between 3D winding (non-trivial) and 2D winding (trivial) spin configurations .
After the training process, the trained machine is scored by a testing dataset with the same composition of nine phases as that in the training (and validation) dataset.
Importantly, without changing the topologies, the Gaussian distributed random transition and random rotation imposed on these three datasets can increase the diversity of dataset and enhance the ability of generalization of the trained machine.

The prediction dataset contains six categories of spin configurations.
The first category is generated with $m$ uniformly distributed from $+3$ to $-3$.
In the second and the third categories, we change the sign of $n_z$ (the second category) and swapping $n_y$ and $n_z$ of $\mathbf{n}$ (the third category).
Finally, we consider three categories for trivial states, which are ferromagnetic (FM), conical, and helical  states. 
FM can be viewed as 1D uncompleted winding configuration while conical and helical can be viewed as 2D uncompleted ones.
In total, we prepared six categories for the prediction dataset.
More details about data preparations will be described in Appendix \ref{app:dataset}.

For the conventional CNN, we use $\mathbf{n}$ as the input data.
For the qCNN, in order to feed the input data into the qCNN classifier, we transform the 3D spin vector into an unit pure quaternion,
\begin{equation}\label{eq:encoding}
    (n_x,n_y,n_z)\in\mathbb{R}^3 \mapsto (0, n_x,n_y,n_z)\in\mathbb{H},
\end{equation}
where the scalar part (the first component) is zero and the vector part is $\mathbf{n}$. Therefore, the inputs of qCNN are effectively equivalent to those of CNN.

\subsection{network structure and performance}

The schematic architectures of these two classifiers are shown in Fig.~\ref{fig:QCNN}, where the last black arrows point to nine neurons for nine topological phases.
In the qCNN classifier, we implement a quaternion convolution (q-Conv) layer as the first layer [red dotted cuboid in Fig.~\ref{fig:QCNN}(b)], and the operations in a q-Conv layer are based on the quaternion algebra to hybridize spin components.
Then the next three layers are typical 3D convolutions (Conv3Ds). Our Conv3Ds do not mix depths by choosing proper sizes of kernels.
Followed the Conv3D layers is a 2D convolution (Conv2D) layer to mix data in depth: nine kernels of kernel size 4×1 will transform data from $4 \times 9$ to $1 \times 9$. On the contrary, the CNN classifer has only Conv2D layers.
Although the qCNN is more complex than the CNN, the total network parameters of the qCNN is however less than the CNN.
This is one advantage of the qCNN over the conventional CNN.

In order for classifiers to satisfy some physically reasonable conditions, two special designs are implemented.
Firstly, we extend the $k$ points out of the BZ by padding the input data according to the periodic boundary conditions~\cite{PhysRevB.99.075113}.
\begin{figure}[b!]
    \centering
    \includegraphics[width=0.45\textwidth]{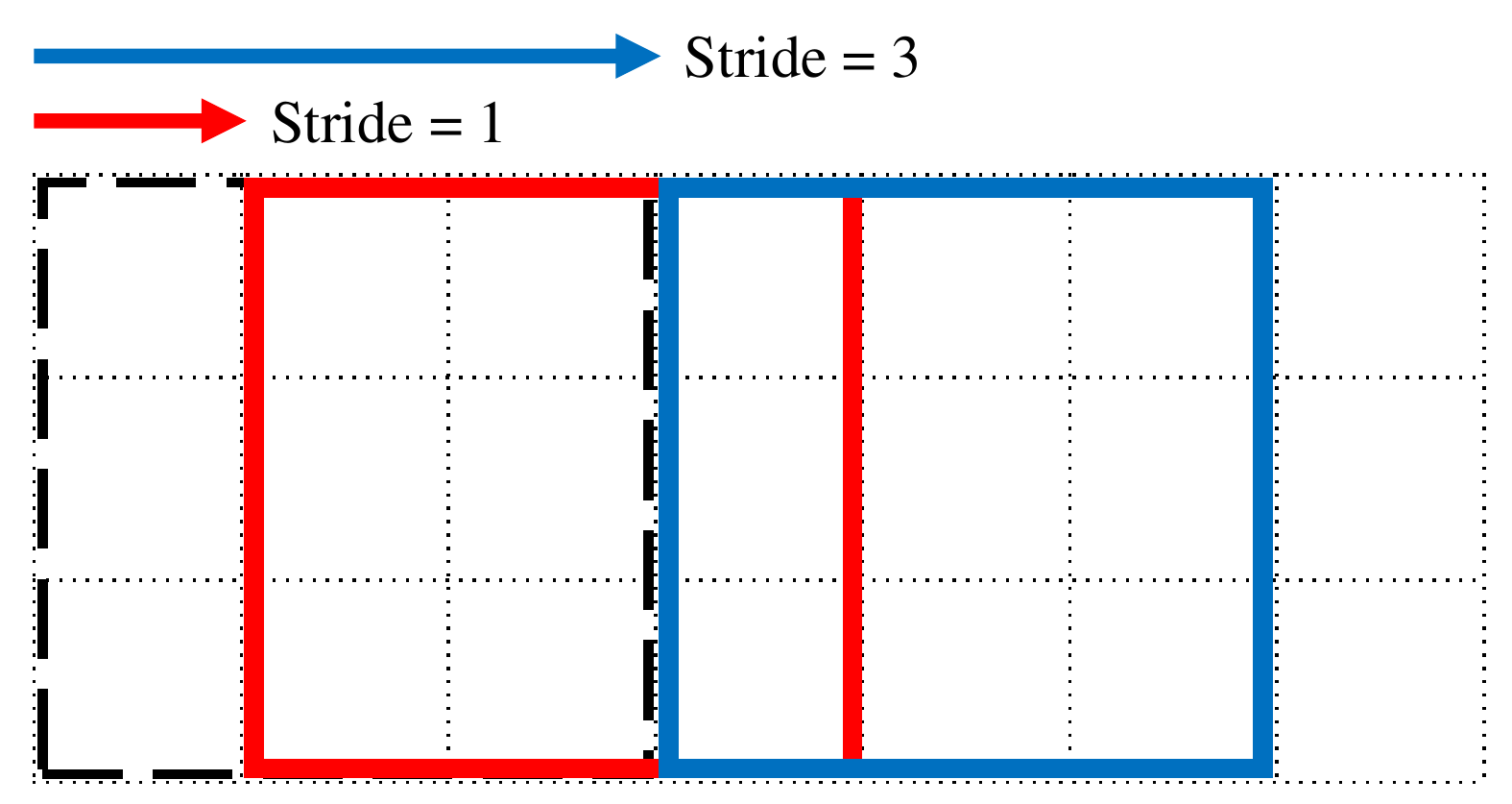}
    \caption{Schematic of ``overlap'' convolution (red solid) and ``non-overlap'' convolution (blue solid) from a $3\times3$ kernel (black dotted) over data. The blue solid square is a signal movement from the kernel, and the size of stride is the same as the length of kernel, thus each movement of this kernel is ``non-overlap.''}
    \label{fig:overlap}
\end{figure}
Secondly, the first layer takes ``overlapping'' strides with an arctan activation function, and the latter layers take ``non-overlapping'' strides with the tanh activation function for both qCNN and CNN machines.
Figure~\ref{fig:overlap} illustrates how the ``overlapping'' and ``non-overlapping'' feature mapping can be manipulated by varying the size of stride.

\begin{figure}[b!]
    \centering
    \includegraphics[width=0.46\textwidth]{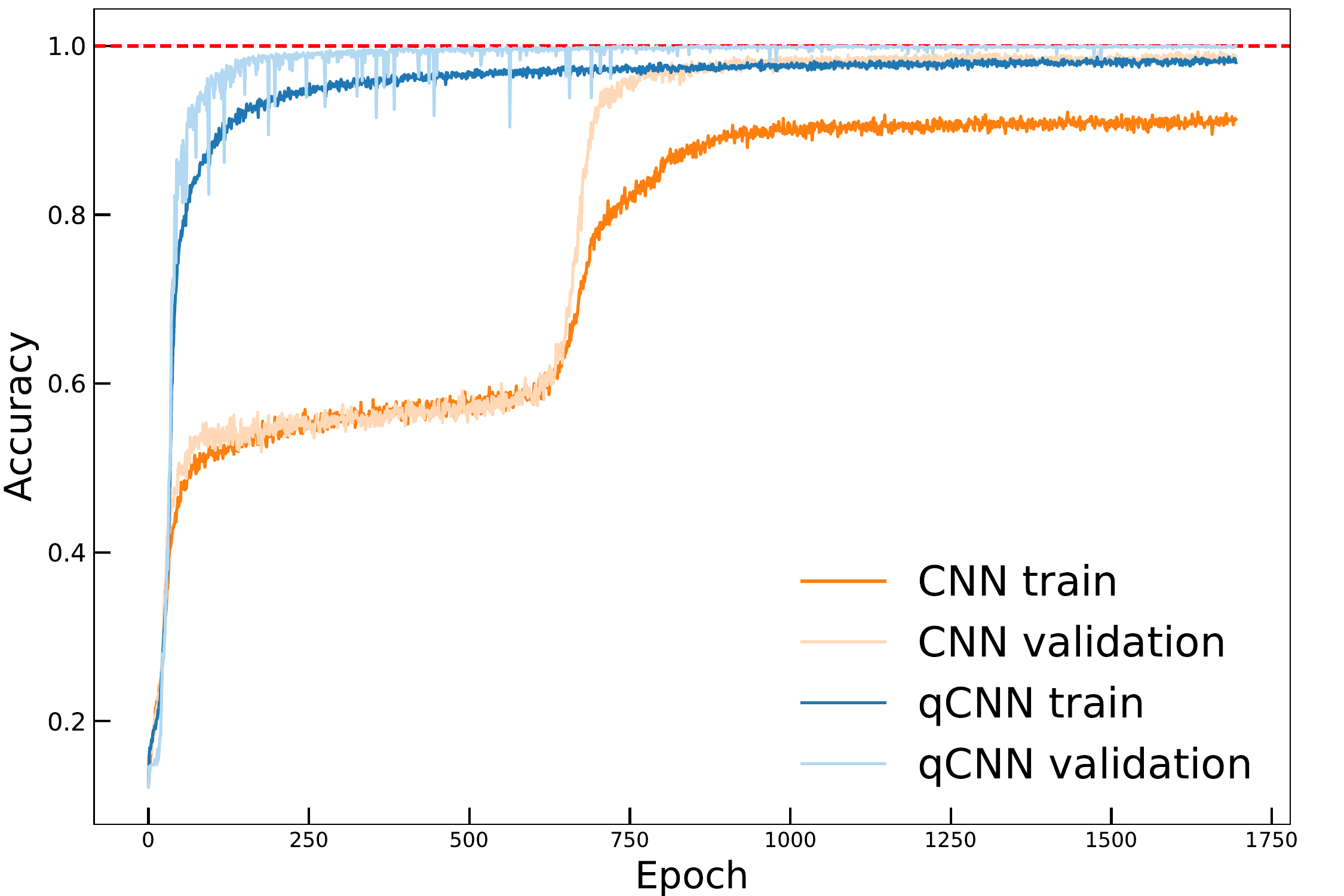}
    \caption{Learning curves of the qCNN and CNN classifiers. By applied \emph{Dropout}, the validation is greater than the training.}
    \label{fig:learningCurve}
\end{figure}

Then, both qCNN and CNN machines are trained.
The learning curves of both machines are shown in Fig.~\ref{fig:learningCurve}.
The CNN machine (orange and light orange lines) jumps over a big learning barrier at around the $700^{th}$ epoch.
After that, the training and the validation accuracy (orange and light orange line respectively) are separated and do not converge up to end of this training process. 
Even though the same training (and validation) dataset is used in the training process, the learning curves of the qCNN machine (blue and light blue lines) are qualitatively different.
The training and the validation accuracy are separated around $90^{th}$ epoch, but the difference between these two accuracies decreases with increasing epochs.
After the training procedure finished, the qCNN (CNN) machine gets 99.67\% (94.12\%) testing accuracy.
This difference in accuracy results from the spin-vortex dataset, where the qCNN works well but CNN dose not.

\begin{figure}[b!]
    \centering
    \includegraphics[width=0.46\textwidth]{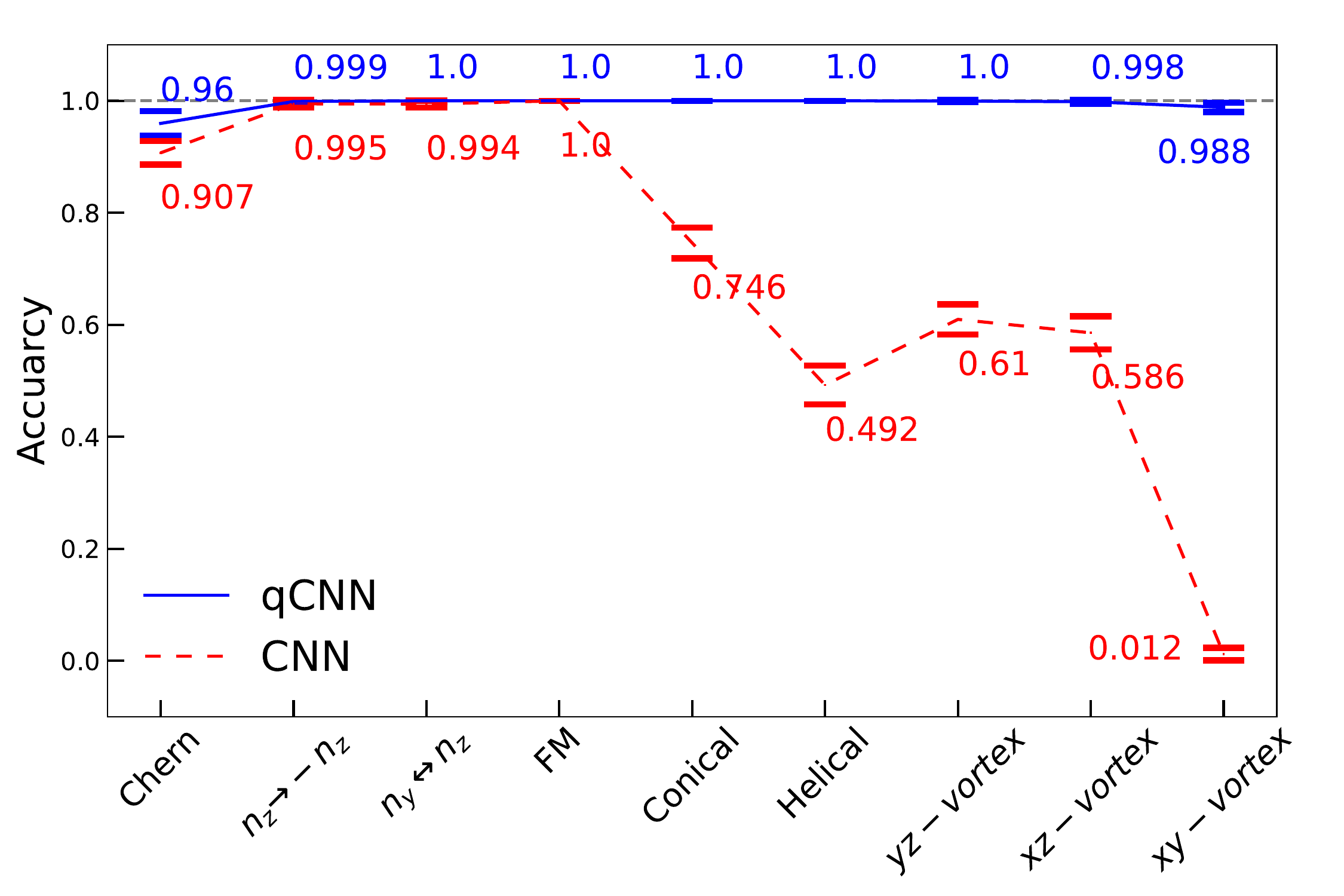}
    \caption{The performance of the qCNN (blue line) and CNN (red dashed line) on the prediction datasets. Numbers tagged are the values of the accuracies. Standard deviations (by error bars) are also provided. The qCNN outperforms the conventional CNN on all prediction datasets, especially on three spin-vortex ones}
    \label{fig:performpred}
\end{figure}

The trained machines are ready to do prediction, and the result is shown in Fig.~\ref{fig:performpred}.
In Fig.~\ref{fig:performpred}, since the first category contains $\mathbf{n}$ of uniformly distributed $m$, where a few data points are very close to the phase boundaries $m \approx \{0, \pm 2 \}$, the accurate rate of the the qCNN is slightly low at $96\%$. For the second and third categories, we choose $m=\pm 1$, away from the phase transition points, and the performance is nearly perfect. For the uncompleted winding configurations, the qCNN, different from the conventional CNN, can accurately classifies FM, helical and conical states after learning the spin-vortex states.
This is the main advantage of the qCNN over the conventional CNN, which is expected to result from the quaternion algebra.

The processing times of two classifiers are summarized in Table~\ref{tab:GPU}.
Since the q-Conv layer has massive matrix multiplication, the time of one epoch for qCNN is longer than that of convention CNN in our task, especially utilized by CPU.

\begin{table}[thb!]
\begin{center}
\begin{tabular*}{0.4\textwidth}{@{\extracolsep{\fill}}c| c c }
\hline
~ & \multicolumn{2}{c}{Processing time (sec)} \\
Architecture & CPU  & GPU  \\
\hline
    CNN  &  6.115  & 1.011   \\
    qCNN  &  72.2   & 3.108\\
    \hline
\end{tabular*}
\caption{Comparison between CPU and GPU processing time (seconds per epoch). Hardware facilities are as follows:
NVIDIA GeForce RTX\textsuperscript{TM} 2080Ti GPU, Intel Xeon\textsuperscript{\textregistered} E5-2650v4 CPU (2.20 GHz Core 12), and 32GB DDR4 SDRAM.}\label{tab:GPU}
\end{center}
\end{table}

\section{discussions}\label{sec:discu}

In this work, we apply the quaternion multiplication laws to both PCA (unsupervised learning) and qCNN (supervised learning). The two methods take different inputs; the former one takes scalar function $F(p)$, which is something to do with a convolution of the wave function, and the second one takes the pure quaternion function $(0,\mathbf{n}(\vec{k}))$, where real part is zero and the imaginary part is the spin vector. We will explain physical intuitions and comment the mechanisms in this section.

On PCA, we did not take $\mathbf{n}$ simply as the input because the representation of the vector $\mathbf{n}$ depends on coordinates but the topology is not. We believed that the topology as a global geometry should be embedded in the correlations. The correlation of dot products of $\mathbf{n}$ turned out to fail since relative angles of two spins were not informative to understand the swirling of $\mathbf{n}$ on $S^2$. If one tries the quaternion in the convolution Eq. (\ref{eq:Ne}) by $q= (0, n_x, n_y, n_z)$, the result is still inappropriate for the convolution is independent of the sign of $m$ to discriminate topological states (see Appendix \ref{app:proof}). Eventually, we found that the $F(p)$ defined in Eq. (\ref{eq:Ne}) was a proper quantity to characterize topology after PCA. The $F$ has the property that it is featured (featureless) when the wave function is unable (able) to be globally continuous that happens in the nontrivial (trivial) phases. Unfortunately, the $F(p)$ is not gauge invariant. The results were based on the choice of gauge in Eqs. (\ref{eq:gauge1}) and (\ref{eq:gauge2}) that made the wave function continuous locally and discontinuous at $k$ where $n_z(\vec{k})=0$. We had examined other choices of gauge and found that the present gauge exhibited the PCA features most clearly (results not shown). We remark that our PCA results looked good because the inputs were ingeniously designed and the PCA method might not be more practical than the qCNN method.

\begin{figure}[b!]
    \centering
    \includegraphics[width=0.47\textwidth]{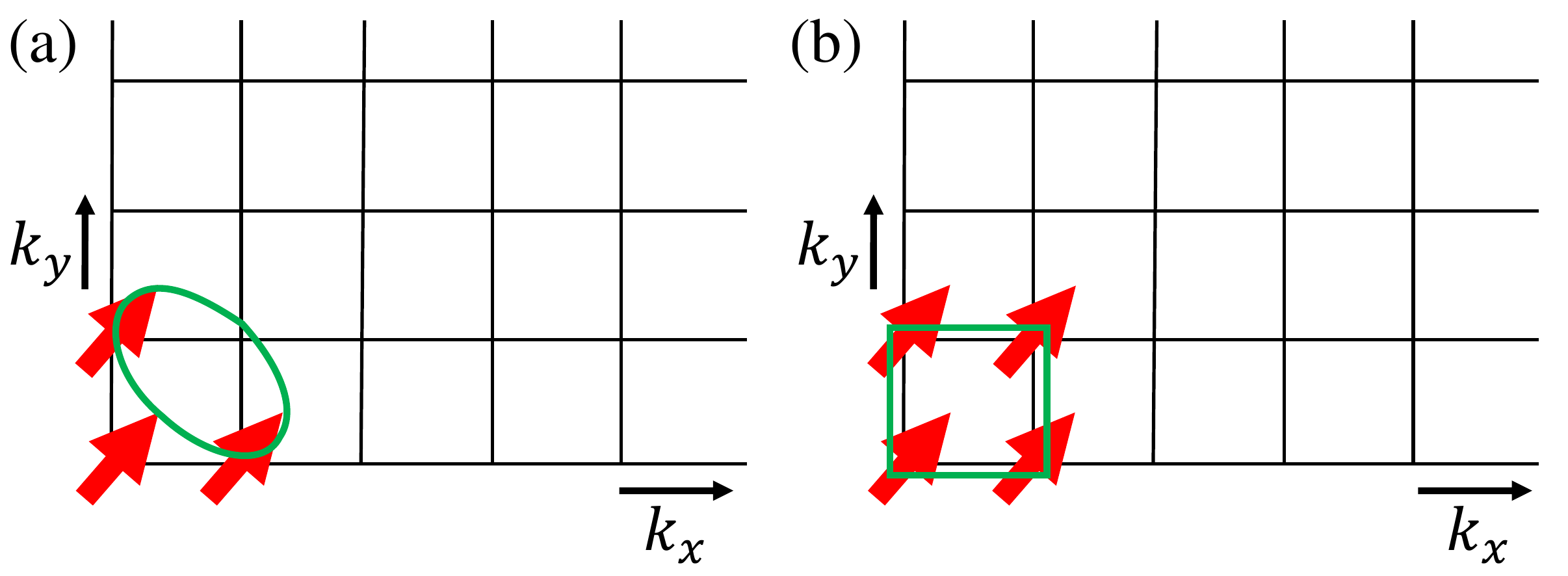}
    \caption{(a) Three nearest neighborhood spin vectors will contribute a solid angle, and (b) four nearest neighbors are enclosed by the kernels in the first convolutional layer.}
    \label{fig:filter}
\end{figure}

\begin{figure}[thb!]
    \centering
    \includegraphics[width=0.49\textwidth]{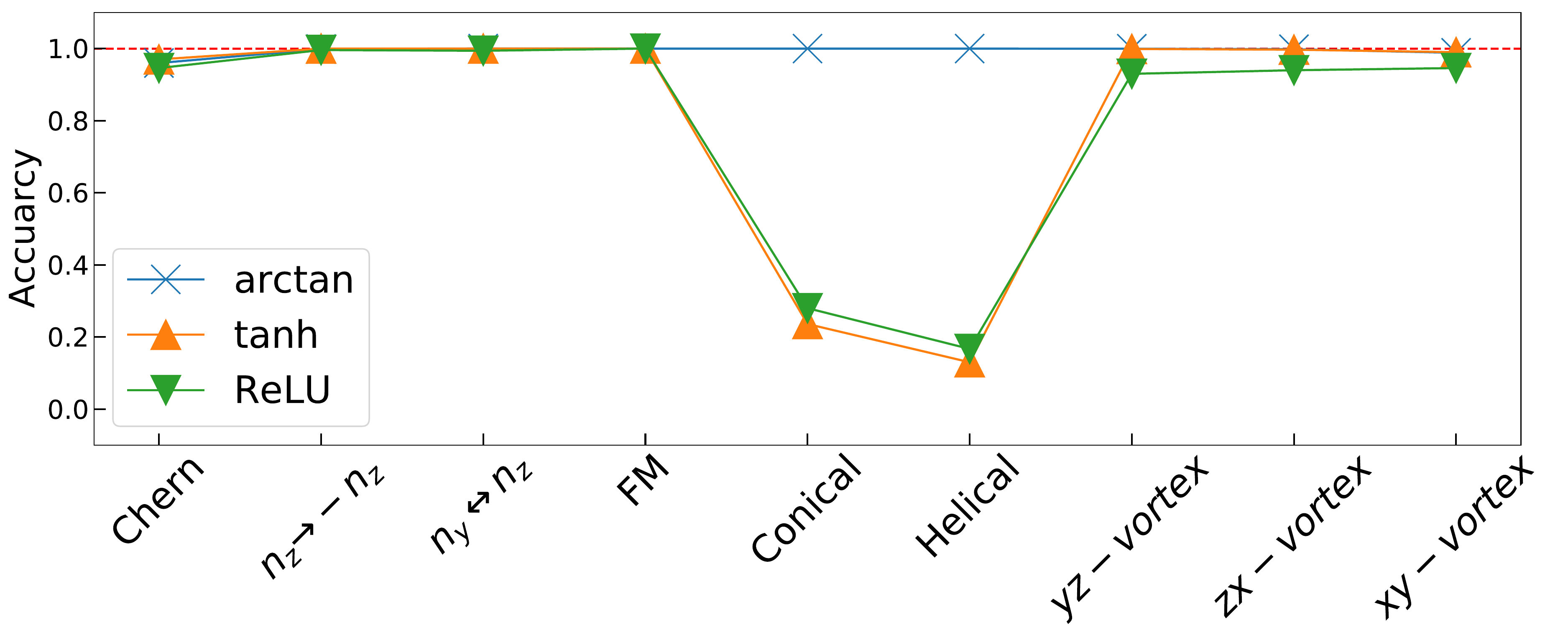}
    \caption{Comparison between three activation functions applied in the first layer of the qCNN classifier.}
    \label{fig:acti}
\end{figure}

For qCNN, it is interesting to understand the mechanism behind. There are several possible factors promoting the performance of our supervised learning machine. The first one is that the size of kernel in the first convolutional layer is 2 × 2 with stride = 1, which means the machine can collect spin information among four nearest neighbors [see Fig. \ref{fig:filter}(b)]. We know that the Chern number is the integral of the Berry curvature in the BZ, and the Berry curvature is twice of the solid angle. A solid angle $\Omega$ subtended by three unit vectors $\vec{a}$, $\vec{b}$, and $\vec{c}$ is obtained by
\begin{equation}\label{eq:solid}
\tan{\frac{\Omega}{2}}=\frac{\abs{\vec{a}\cdot(\vec{b}\times \vec{c})}}{1+\vec{a}\cdot \vec{b}+\vec{b}\cdot \vec{c}+\vec{c}\cdot \vec{a}}.
\end{equation}
Our choice of the size of the kernel in the first hidden layer is the minimal of $2 \times 2$ that mixes only the nearest-neighboring spins. In this way, it is very possible to enforce the machine to notice the solid angle extended in this plaquette. The second factor is the quaternion product. Recall that the conventional CNN might correlate spins $\mathbf{n}'s$ in neighboring $\vec{k}'s$ due to the feature map through the kernel. However, the map does not mix the components of spins. In comparison, the qCNN is more efficient for it directly entangle spins via the quaternion product. It is this entanglement of spin components by the quaternion product that makes the scalar and vector products in calculating the solid angle (see Eq.~ (\ref{eq:solid})) become possible to be realized by the machine. As a solid angle involves at least three spins and the feature map by the kernel is just linear, a nonlinear transformation is crucial to create high-order (three spins) terms in the expansion. This is possible and proved in Ref. \cite{lin2017} that multiplication of variables can be accurately realized by simple neural nets with some smooth nonlinear activation function. Therefore, the third factor is the non-linear activation function, arctan in this work. We expect that using arctan as the activation function can further help the machine to learn correct representations because the calculation of a solid angle involves the arctan operation in Eq. (\ref{eq:solid}). This belief is indeed supported by the results shown in Fig. \ref{fig:acti}, where the arctan activation function outperforms the ReLU and tanh activation functions over nine different datasets. In summary, several factors are combined to enhance the performance of our machine as follows. The quaternion-based operations in the q-Conv layer mix not only spins with their neighbors but also components of spins. When these linear combinations are fed into the non-linear activation functions in our qCNN, the output can be viewed as an expansion of a non-linear function, which may contain a term having both the scalar- and vector-product of neighboring spins, similar to that in Eq.~ (\ref{eq:solid}). Therefore, after the optimization process, the machine may keep increasing the weight of a solid-angle-related term and eventually learn to classify the topological phases.

Also, adding some noises to the training dataset helped our supervised-learning machine to learn the generic feature of our data.
We found that when the training data was generated directly from Eq. (\ref{eq:HighChern}) without adding any noise, the machine worked well for training and testing datasets but had poor performance on all the prediction dataset. This could be understood by noting that the topological invariant is determined by the sign of $m$, which appears in the $z$ component in Eq. (\ref{eq:HighChern}). By using the dataset without noise, the machine might naively regard the $z$ component as the judgment of topology when the training data does not contain wide distribution. We note that the topology is invariant when the spin texture is uniformly translated or rotated. So we trained our machine with randomly translated and rotated data to avoid incorrect learning. (See data preparation in Appendix \ref{app:dataset}.) From our observations, the performance on the prediction dataset was remarkably enhanced when the noise was included, which supports our idea.

\section{Conclusions}
In summary, we classify topological phases with distinct Chern numbers via two types of machine-learning techniques.
For the unsupervised part, we propose a quaternion-based convolution to transform the topological states into the input data. With this convolution, distinct topological states are successfully classified by PCA, a linear machine for classification. 

We then go to the supervised learning part where, in contrast to the conventional CNN, we successfully use the qCNN to classify different topological phases.
This work demonstrates the power of quaternion-based algorithm, especially for the topological systems with the Chern number as the topological invariants.

\begin{acknowledgments}
This study is supported by the Ministry of Science and Technology (MoST) in Taiwan under grant No. 108-2112-M-110-013-MY3. M.R.L. and W.J.L. contributed equally to this work.
\end{acknowledgments}

\appendix
\section{Data preparation} \label{app:dataset}
\noindent\emph{Training dataset}---
The normalized spin configurations $\mathbf{n}^{(c,m)}(\vec{k}),\forall\vec{k}\in\mathrm{BZ}$ are based on the formula [refer to Eq.~(\ref{eq:HighChern})] 
\begin{equation*}
\begin{split}
\mathbf{n}^{(c,m)}&(\vec{k}):=\mathbf{h}^{(c,m)}(\vec{k})/\norm{\mathbf{h}^{(c,m)}(\vec{k})},\quad \textrm{where}\\
&\mathbf{h}^{(c,m)}(\vec{k}) = 
    \begin{pmatrix}
        \ \mathrm{Re}\big[(\sin{k_x}-i\sin{k_y})^c\big]\\
        -\mathrm{Im}\big[(\sin{k_x}-i\sin{k_y})^c\big]\\
        \cos{k_x}+\cos{k_y}+m
    \end{pmatrix}
\end{split}\end{equation*}
in a $40\times 40$ square lattice with periodic boundary conditions. For each $c=1,2,3$ and $4$, we generated four sets $S_1^{(c)}, S_2^{(c)}$, $S_3^{(c)}$ and $S_4^{(c)}$. The former two sets are topologically nontrivial and each has 640 configurations for different values of $m$:
\begin{equation*}\begin{matrix}
S_1^{(c)} =& 
\big\{
    \mathbf{n}^{(c,m)}(\vec{k}):m\in[-1.9,-0.1],~\vec{k}\in\mathrm{BZ}
\big\},\\
S_2^{(c)} =& 
\big\{
    \mathbf{n}^{(c,m)}(\vec{k}):m\in[0.1,1.9],~\vec{k}\in\mathrm{BZ}
\big\},
\end{matrix}
\end{equation*}
where $m$ are random numbers in the corresponding ranges. The latter two sets are topologically trivial and each has 80 (identical) configurations:
\begin{equation*}
    \begin{matrix}
        S_3^{(c)} =& \big\{\mathbf{n}^{(c,m)}(\vec{k}):m=-3,~\vec{k}\in\mathrm{BZ}\big\},\\
        S_4^{(c)} =& \big\{\mathbf{n}^{(c,m)}(\vec{k}):m=3,~\vec{k}\in\mathrm{BZ}\big\}.
    \end{matrix}
\end{equation*}
So, for each $c$ there were 1280 nontrivial spin configurations and 160 trivial ones. Then the primitive data passed through some manipulations as the effect of data augmentation without changing the topologies. Each spin configuration $\mathbf{n}(\vec{k})$ was translated ($\mathcal{T}$), rotated ($\mathcal{R}$), and then polluted with noise ($\mathcal{G}$):
\begin{equation}\label{eq:arg}
   \mathbf{n}(\vec{k})\xrightarrow{\mathcal{T}}\mathbf{n}(\vec{k}+\vec{p}_0)\xrightarrow{\mathcal{R}}\mathbf{n}'(\vec{k}+\vec{p}_0)\xrightarrow{\mathcal{G}}\mathbf{n}'(\vec{k}+\vec{p}_0)+\Delta\mathbf{n}'(\vec{k}),
\end{equation}
where $\vec{p}_0$ is a random displacement in $\vec{k}$, $\mathcal{R}$ stands for a random 3D rotation of the 
spin, and $\Delta\mathbf{n}'(\vec{k})$ is Gaussian noise ($\mathcal{G}$) with standard deviation $0.1\pi$ in each component. (The spin should be normalized lastly.) $\mathcal{T}$ and $\mathcal{R}$ are homogeneous transformations in $\vec{k}$, but $\mathcal{G}$, inhomogeneous, picks only 30 out of 1600 $\vec{k}$ sites. 

In addition to the 5760 sets of data in nine topological phases ($C=-4$ to $C=+4$), we also include 360 spin vortex states, which are $C=0$ states, based on the formulas:
\begin{equation}\label{vortex1}
{}^{yz}\mathbf{h}^{(c,m)}
(\vec{k})=
    \begin{pmatrix}
        0\\
        -\mathrm{Im}\big[(\sin{k_x}-i\sin{k_y})^c\big]\\
        \cos{k_x}+\cos{k_y}+m
    \end{pmatrix}
\end{equation}
\begin{equation}\label{vortex2}
{}^{xz}\mathbf{h}^{(c,m)}(\vec{k})=
    \begin{pmatrix}
        \ \mathrm{Re}\big[(\sin{k_x}-i\sin{k_y})^c\big]\\
        0\\
        \cos{k_x}+\cos{k_y}+m
    \end{pmatrix}
\end{equation}
\begin{equation}\label{vortex3}
{}^{xy}\mathbf{h}^{(c)}(\vec{k})=
    \begin{pmatrix}
        \ \mathrm{Re}\big[(\sin{k_x}-i\sin{k_y})^c\big]\\
        -\mathrm{Im}\big[(\sin{k_x}-i\sin{k_y})^c\big]\\
        0
    \end{pmatrix}    
\end{equation}
with their normalized configurations.
For each $c$, 30 spin configurations were generated with random $m$ ranging from -3 to 3. The data also went through translation $\mathcal{T}$ and rotation $\mathcal{R}$ but no noise $\mathcal{G}$.

Therefore, we generated 6120 spin configurations totally as the training dataset.
Among the training dataset, 25\% of the data are assigned as the validation dataset (light color lines in Fig.~\ref{fig:learningCurve}). 
\\\\
\noindent\emph{Testing dataset}---
In addition to the training and validation dataset, we prepare extra 1224 spin configurations as the testing dataset, with the same composition as the training and validation datasets. This dataset is prepared for scoring the trained classifiers.
\\\\
\begin{figure}[h!]
    \centering
    \includegraphics[width=0.47\textwidth]{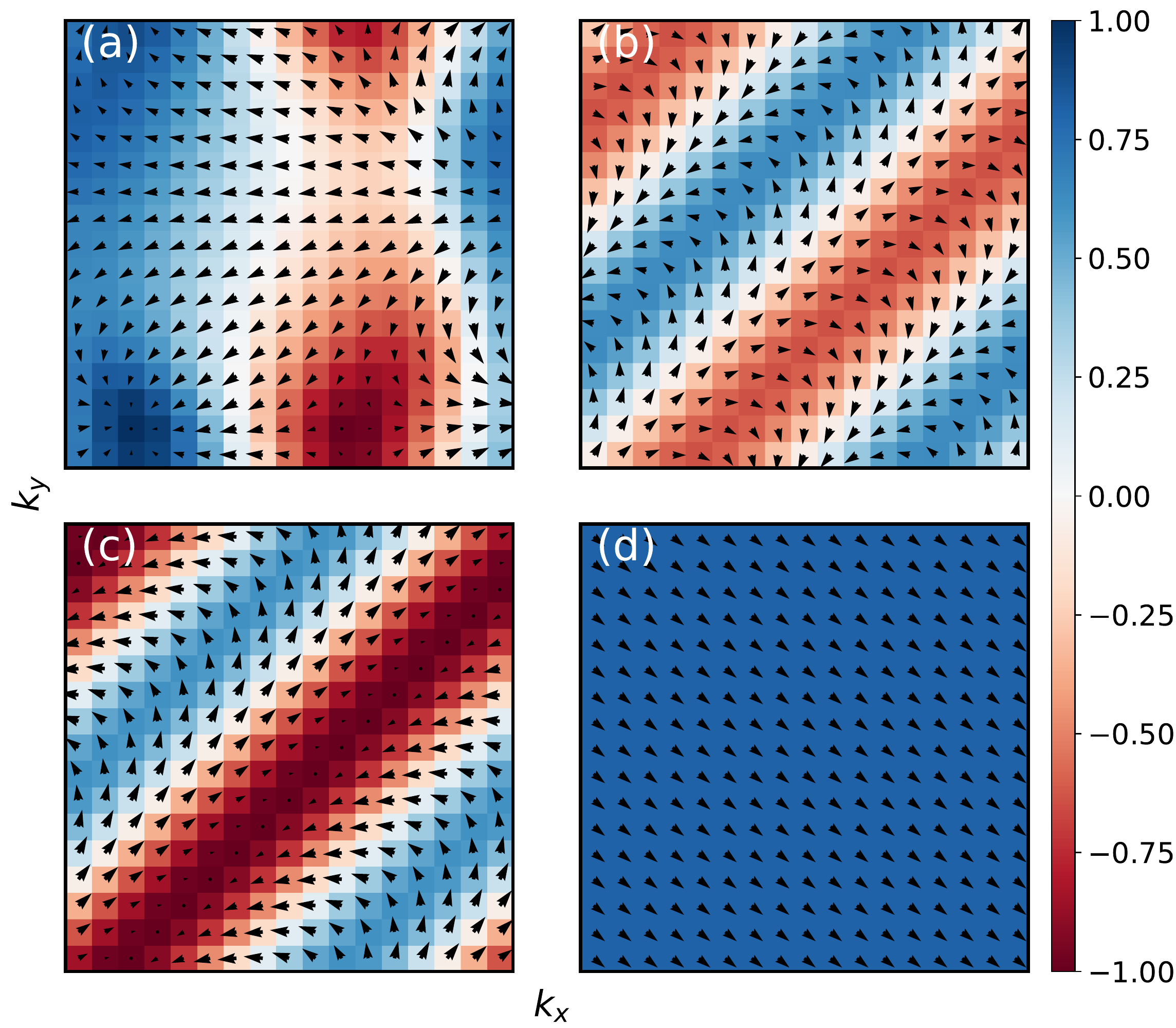}
    \caption{Some examples in the prediction dataset for states: (a) $n_y\leftrightarrow n_z$, (b) helical, (c) conical, and (d) FM.}
    \label{fig:trivial}
\end{figure}

\noindent\emph{Prediction dataset}---
The prediction dataset is an extra dataset, different from the aforementioned three datasets.
It consists of six categories, each of which was not seen by the machine during the training process.
This dataset was processed by $\mathcal{T}$ and $\mathcal{R}$ but not $\mathcal{G}$. The six categories were constructed as following.
The First category, the ``chern'' category, is a set $S$ which was generated from Eq.~(\ref{eq:HighChern}) with 30 $m$'s uniformly ranging from -3 to 3 for each $c$:
$$
S = \big\{\mathbf{n}^{(c)}:c= 1,2,3,4,~m=-3+\frac{6i}{29},~i = 0,...,29\big\}
$$
As a reminder, this category is different from the training dataset.
The training data includes the specific $m=\pm3$ at trivial phase, and two intervals $\{\left[-1.9,-0.1\right]$ and $\left[0.1,1.9\right]\}$ in nontrivial phases.
Therefore, 20\% of this category is close to the phase transition $m\approx\{0,+2,-2\}$.

The next two categories were generated based on Eq.~(\ref{eq:HighChern}) with $m = \pm1$.
The first one was constructed by changing the sign of the $z$-component:
\begin{equation*}
    \begin{split}
    \mathbf{h}(\vec{k}) = 
        \begin{pmatrix}
            \ \mathrm{Re}\big[(\sin{k_x}-i\sin{k_y})^c\big]\\
            -\mathrm{Im}\big[(\sin{k_x}-i\sin{k_y})^c\big]\\
            \cos{k_x}+\cos{k_y}+m
        \end{pmatrix}\\
    \to 
        \begin{pmatrix}
            \ \mathrm{Re}\big[(\sin{k_x}-i\sin{k_y})^c\big]\\
            -\mathrm{Im}\big[(\sin{k_x}-i\sin{k_y})^c\big]\\
            - \cos{k_x} -\cos{k_y} - m
        \end{pmatrix}.
    \end{split}
\end{equation*} 
The second one was constructed by swapping the $y$ and the $z$ components:
\begin{equation*}
    \begin{split}
        \mathbf{h}(\vec{k}) = 
            \begin{pmatrix}
                \ \mathrm{Re}\big[(\sin{k_x}-i\sin{k_y})^c\big]\\
                -\mathrm{Im}\big[(\sin{k_x}-i\sin{k_y})^c\big]\\
                \cos{k_x}+\cos{k_y}+m
            \end{pmatrix}\\
        \to 
            \begin{pmatrix}
                \ \mathrm{Re}\big[(\sin{k_x}-i\sin{k_y})^c\big]\\
                \cos{k_x}+\cos{k_y}+m\\
                -\mathrm{Im}\big[(\sin{k_x}-i\sin{k_y})^c\big]
            \end{pmatrix}.
    \end{split}
\end{equation*}

The next two categories, called helical and conical spin configurations, were generated based on the following equation
$$
\mathbf{n}_{\mathrm{helical/conical}}(\vec{k})=
    \begin{pmatrix}
        \sqrt{1-\epsilon^2}\cos{(k_x+ k_y)}\\
        \sqrt{1-\epsilon^2}\sin{(k_x+ k_y)}\\
        \epsilon
    \end{pmatrix}.
$$
Here $\epsilon =0 $ is for the helical state and $0<|\epsilon|<1$ is for a conical state. 
The last category contains the ferromagnetic spin configurations (FM) whose $z$-component are a constant and $x$- and $y$-component are zero. Some spin configurations in the prediction dataset are illustrated in Fig.~\ref{fig:trivial}.

\section{Quaternion} \label{app:qcnn}
The quaternion number system were introduced by Irish mathematician William Rowan Hamilton in 1843 as an extension of the complex numbers. 
A quaternion number $q$ is composed of four real numbers \(r,~a,~b\) and \(c\) to be 
\begin{equation}
   q = r+a\hat{\mathbf{i}}+b\hat{\mathbf{j}}+c\hat{\mathbf{k}}, 
\end{equation}
where \( \{ \pmb{1},~\hat{\mathbf{i}},~\hat{\mathbf{j}},~ \hat{\mathbf{k}} \} \) is the basis. 
Sometimes it is written as \(q= (r,\ \vec{v}) \) or \(q= (r,a,b,c) \) in short. 
Here \(r\) is called the scalar (or real) part of the quaternion and \(\vec{v} = (a,b,c)\) the vector (or imaginary) part. 
A quaternion without scalar part \(q = (0, a, b, c)\) is called \emph{pure quaternion}.
Similar to the imaginary number, 
\begin{equation} \label{alg1}
    \hat{\mathbf{i}}^2 = \hat{\mathbf{j}}^2 = \hat{\mathbf{k}}^2 = \hat{\mathbf{i}} \hat{\mathbf{j}} \hat{\mathbf{k}} = -\pmb{1}.
\end{equation}
Importantly, the algebra of quaternions is noncommutative, based on
\begin{gather} \label{alg2}
    \pmb{1}\hat{\mathbf{i}} =\hat{\mathbf{i}}\pmb{1}=\hat{\mathbf{i}},\quad 
    \pmb{1}\hat{\mathbf{j}} = \hat{\mathbf{j}}\pmb{1} = \hat{\mathbf{j}}, \quad
    \pmb{1}\hat{\mathbf{k}}= \hat{\mathbf{k}}\pmb{1}= \hat{\mathbf{k}}, \\\nonumber
    \hat{\mathbf{i}}\hat{\mathbf{j}} = -\hat{\mathbf{j}}\hat{\mathbf{i}} = \hat{\mathbf{k}},\quad
    \hat{\mathbf{j}}\hat{\mathbf{k}} = -\hat{\mathbf{k}}\hat{\mathbf{j}} = \hat{\mathbf{i}},\quad\mathrm{and}\quad
    \hat{\mathbf{k}}\hat{\mathbf{i}} = -\hat{\mathbf{i}}\hat{\mathbf{k}} = \hat{\mathbf{j}}.\nonumber
\end{gather}
The conjugate of the quaternion is defined to be 
\begin{equation}
  q^{*} = r-a\hat{\mathbf{i}}-b\hat{\mathbf{j}}-c\hat{\mathbf{k}},  
\end{equation}
and the norm is given by
\begin{equation}
    ||q||=\sqrt{q q^{*}} = \sqrt{r^2 + a^2 +b^2 +c^2 }.
\end{equation}
Therefore the inverse of $q$ is defined as
\begin{equation}
    q^{-1}:=\frac{q^*}{\norm{q}^2}.
\end{equation}
If \(q\) is unit quaternion, then their inverse is exactly their conjugate.
The multiplication (so-called quaternion product or Hamilton product) of two quaternions \(q_1 = (r_1,~a_1,~b_1,~c_1)\) and \(q_2 = (r_2,~a_2,~b_2,~c_2)\) is given by
\begin{equation} \label{q1q2}
    \begin{split}
    q_1 q_2 & = (r_1 r_2 - a_1 a_2 - b_1 b_2 - c_1 c_2) \\
      & \quad + (a_1 r_2 + r_1 a_2 - c_1 b_2 + b_1 c_2)\hat{\mathbf{i}} \\
      & \quad + (b_1 r_2 + c_1 a_2 + r_1 b_2 - a_1 c_2)\hat{\mathbf{j}} \\
      & \quad + (c_1 r_2 - b_1 a_2 + a_1 b_2 + r_1 c_2)\hat{\mathbf{k}} \\
    &= (r_1 r_2 -\vec{v}_1 \cdot \vec{v}_2,~ r_1 \vec{v}_2 + r_2 \vec{v}_1 + \vec{v}_1\times \vec{v}_2).
\end{split}
\end{equation}

To realize the algebra in Eqs.~(\ref{alg1}) and (\ref{alg2}), one can choose the \(\mathcal{M}(4, \mathbb{R})\) representation for the quaternion numbers with
\[
\begin{split}
    \pmb{1} \doteq & 
        \begin{pmatrix}
            1 & 0 & 0 & 0 \\
            0 & 1 & 0 & 0 \\
            0 & 0 & 1 & 0 \\
            0 & 0 & 0 & 1 
        \end{pmatrix},\ 
    \hat{\mathbf{i}} \doteq
        \begin{pmatrix*}[r]
            0 & -1 & 0 & 0 \\
            1 & 0 & 0 & 0 \\
            0 & 0 & 0 & -1 \\
            0 & 0 & \ 1 & 0 
        \end{pmatrix*},\\
    &\hat{\mathbf{j}} \doteq
        \begin{pmatrix*}[r]
            0 & 0 & -1 & 0 \\
            0 & 0 & 0 & \ 1 \\
            1 & 0 & 0 & 0 \\
            0 & -1 & 0 & 0 
        \end{pmatrix*},\ \mathrm{and}\ 
    \hat{\mathbf{k}} \doteq
        \begin{pmatrix*}[r]
            0 & 0 & 0 & -1 \\
            0 & 0 & -1 & 0 \\
            0 & \ 1 & 0 & 0 \\
            1 & 0 & 0 & 0 
        \end{pmatrix*},
\end{split}
\]
so that
\[
q \doteq
\begin{pmatrix*}[r]
    r & -a & -b & -c\\
    a & r & -c & b\\
    b & c & r & -a\\
    c & -b & a & r
\end{pmatrix*}.
\]
Reversely, 
\[
r=\frac{1}{4}\mathrm{tr}(q),~ a=-\frac{1}{4} \mathrm{tr}(\hat{\mathbf{i}} q),~
b=-\frac{1}{4}\mathrm{tr}(\hat{\mathbf{j}} q),~c=-\frac{1}{4}\mathrm{tr}(\hat{\mathbf{k}} q).
\]
It is evident that in terms of matrices the commutativity of multiplication of quaternions dose not hold. 
Furthermore, in the matrix representation \(q^*=q^T\), conjugation of a quaternion being equal to its transposition.
More specifically, an unit quaternion have a property $q^{-1} = q^* = q^{T}$ in the $\mathcal{M}(4,\mathbb{R})$ representations.

\section{Details of definition in Section III} \label{app:proof}
In this section, we provide some properties about the $F(\vec{p})$ function we defined in PCA section and the convolution of normalized spin vector $\mathbf{n}$.
Recall that the convolution we defined in the PCA section is as follows
\begin{equation}
\begin{split}
    F(\vec{p}) :=  (\overline{q}^*&\circledast\overline{q})[\vec{p}] - (\underline{q}^*\circledast\underline{q})[\vec{p}]
    \quad\mathrm{with}\\
    (q^*&\circledast q)[\vec{p}] :=  \sum_{\vec{k}\in \mathrm{BZ}} q^*\bigg\rvert_{\vec{k}}q\bigg\rvert_{\vec{p} - \vec{k}}.
\end{split}
\end{equation}
From now on, because of lack of notations, we denoted upper-bar (e.g.:$~\overline{q},\ket{\overline{u}},\bra{\overline{u}},\overline{h}$)  being conduction band, and lower-bar (e.g.~$\underline{q},\ket{\underline{u}},\bra{\underline{u}},\underline{h}$) being valence band.
A vertical line with a variable stand for its corresponding position of BZ. 
\begin{Pro}\label{thm1}
    $F(\vec{p})$ is a purely real-valued function.
\end{Pro}
\begin{proof}
Since $\vec{p}-\vec{k}$ and $\vec{k}$ are one-to-one correspondence in BZ, and summing over $\vec{k}\in$BZ or $\vec{p}-\vec{k}\in$BZ are equivalent. Once we take conjugate on $F$, then
\begin{equation*}
\begin{split}
F^*(\vec{p})
&=\sum_{\vec{k}\in\mathrm{ BZ}}\overline{q}^*\bigg\rvert_{\vec{p}-\vec{k}} \overline{q}\bigg\rvert_{\vec{k}}-\underline{q}^*\bigg\rvert_{\vec{p}-\vec{k}}\underline{q}\bigg\rvert_{\vec{k}}\\
&=\sum_{\vec{p}-\vec{k}\in \mathrm{BZ}}\overline{q}^*\bigg\rvert_{\vec{p}-\vec{k}} \overline{q}\bigg\rvert_{\vec{k}}-\underline{q}^*\bigg\rvert_{\vec{p}-\vec{k}}\underline{q}\bigg\rvert_{\vec{k}}\\
&=\sum_{\vec{k}'\in\mathrm{BZ}}\overline{q}^*\bigg\rvert_{\vec{k}'} \overline{q}\bigg\rvert_{\vec{p}-\vec{k}'}-\underline{q}^*\bigg\rvert_{\vec{k}'}\underline{q}\bigg\rvert_{\vec{p}-\vec{k}'}\\
&=F(\vec{p}).
\end{split}
\end{equation*}
The first line come from the property of conjugate on quaternions, the second line come from the equivalence of summing whole BZ, and the third line come from $\vec{k}'\leftrightarrow\vec{p}-\vec{k}$.
We see that conjugation of $F$ is itself.
Therefore, $F(\vec{p})$ is a purely real-valued function.
\end{proof}

Recall that in our model Eq.~(\ref{eq:HighChern}), $h_1,~h_2$ are both even of $\vec{k}$ and $(h_3-m)$ is odd of $\vec{k}$. 
That is, given $\vec{k}=(k_x,k_y)\in\mathrm{BZ}$, there is $\vec{k}'=(\pi-k_x,~\pi-k_y)\in\mathrm{BZ}$ such that
\begin{equation}\label{eq:app_symm}
\left\{
    \begin{matrix}
    h_1\underset{\vec{k}}{\big\rvert} &=& h_1\underset{\vec{k}'}{\big\rvert}\\
    h_2\underset{\vec{k}}{\big\rvert} &=& h_2\underset{\vec{k}'}{\big\rvert}\\
    h_3(m)\underset{\vec{k}}{\big\rvert} - m &=& -\Big(h_3(m')\underset{\vec{k}'}{\big\rvert} - m'\Big)
    \end{matrix}.
\right.
\end{equation}
In addition, those two points $\vec{k},\vec{k}'$ are one-to-one correspondence in BZ, and identical at the $\Gamma$ point.
Notice that once we normalized $\mathbf{h}(m)=(h_1,h_2,h_3(m))$ by $\norm{\mathbf{h}(m)}$, then each component of $\mathbf{n}(m)$ is function of $m$.
\begin{Pro}\label{thm2}
Encoding $\mathbf{n}(m)$ into quaternion by $q=(0,n_x(m),n_y(m),n_z(m))$, the convolution $q^*\circledast q$ is independent on the sign of $m$.
\end{Pro}
\begin{proof}
We consider two convolutions with $q=(0,~\mathbf{n}(m))$ but over $\vec{k}$, $\vec{k}'=(\pi-k_x,\pi-k_y)$ and opposite sign $m'--m$, respectively:
\begin{equation}\label{eq:app_LHS}
\begin{aligned}[b]
(q^*\circledast q)[\vec{p},m] 
&=\sum_{\vec{k}\in\mathrm{BZ}}\big(0,-\mathbf{n}(m)\big)\underset{\vec{k}}{\Big\rvert}\big(0,~\mathbf{n}(m)\big)\underset{\vec{p}-\vec{k}}{\Big\rvert}\\
&=\sum_{\vec{k}\in\mathrm{BZ}}\Big(
n_x(m)\underset{\vec{k}}{\big\rvert}n_x(m)\underset{\vec{p}-\vec{k}}{\big\rvert}\\
&\hspace{3em} +
n_y(m)\underset{\vec{k}}{\big\rvert}n_y(m)\underset{\vec{p}-\vec{k}}{\big\rvert}\\
&\hspace{3em} +
n_z(m)\underset{\vec{k}}{\big\rvert}n_z(m)\underset{\vec{p}-\vec{k}}{\big\rvert},~\vec{V}_L\underset{\vec{k}}{\big\rvert}
\Big),
\end{aligned}
\end{equation}
and
\begin{equation}\label{eq:app_RHS}
\begin{aligned}[b]
(q^*\circledast q)[\vec{p},-m]&=\sum_{\vec{k}'\in\mathrm{BZ}}\big(0,-\mathbf{n}(-m)\big)\underset{\vec{k}'}{\Big\rvert}\big(0,~\mathbf{n}(-m)\big)\underset{\vec{p}-\vec{k}'}{\Big\rvert}\\
&=\sum_{\vec{k}'\in\mathrm{BZ}}\Big(
n_x(-m)\underset{\vec{k}'}{\big\rvert}n_x(-m)\underset{\vec{p}-\vec{k}'}{\big\rvert}\\
&\hspace{3.5em} +
n_y(-m)\underset{\vec{k}'}{\big\rvert}n_y(-m)\underset{\vec{p}-\vec{k}'}{\big\rvert}\\
&\hspace{3.5em} +
n_z(-m)\underset{\vec{k}'}{\big\rvert}n_z(-m)\underset{\vec{p}-\vec{k}'}{\big\rvert},\vec{V}_R\underset{\vec{k}'}{\big\rvert}
\Big)
\end{aligned}
\end{equation}
where $\vec{V}_L,\vec{V}_R$ are vector parts of the above quaternion product at $\vec{k},\vec{k}'$, respectively.
In Property~(\ref{thm1}), we have shown the convolution over entire BZ is a purely real-valued function.
That is, we only need to consider dot product of vector part as the quaternion product of two quaternion $q_1, q_2$ when $q_1,q_2$ both doesn't have real part.
Now, the Eq.~(\ref{eq:app_symm}) and the assumption $m'=-m$ gives us the fact
\begin{gather}
    h_3(m)\underset{\vec{k}}{\Big\rvert}-m = - (h_3(m')\underset{\vec{k}'}{\Big\rvert}-m')=-h_3(-m)\underset{\vec{k}'}{\Big\rvert}-m. \notag \\
    \Rightarrow h_3(m)\underset{\vec{k}}{\Big\rvert} = - h_3(-m)\underset{\vec{k}'}{\Big\rvert}\quad\forall\vec{k},\vec{k}'\in\mathrm{BZ}. \notag
\end{gather}
Since $\norm{\mathbf{h}(\vec{k},m)} = \norm{\mathbf{h}(\vec{k}',-m)}$, we can conclude that
\begin{equation}\label{eq:app_symm2}
\left\{
    \begin{matrix}
    n_x(m)\underset{\vec{k}}{\big\rvert} &=& n_x(-m)\underset{\vec{k}'}{\big\rvert}\\
    n_y(m)\underset{\vec{k}}{\big\rvert} &=& n_y(-m)\underset{\vec{k}'}{\big\rvert}\\
    n_z(m)\underset{\vec{k}}{\big\rvert} &=&
    -n_z(-m)\underset{\vec{k}'}{\big\rvert}
    \end{matrix},
\quad\forall\vec{k},\vec{k}'\in\mathrm{BZ}\right.
\end{equation}
Substituting Eq.~(\ref{eq:app_symm2}) into Eq,~(\ref{eq:app_RHS}), we identify Eqs.~(\ref{eq:app_LHS}) and (\ref{eq:app_RHS}). 
Therefore, applying opposite $m$, the convolutions over BZ gets exactly the same value.
That is, the convolution $q^*\circledast q$ is independent on the sign of $m$ once we encoded quaternion by $(0,~\mathbf{n}(m))$.
\end{proof}

Property~(\ref{thm2}) is based on how we encoded quaterion number.
The following two properties are based on the way we transformed quaternion in main text.

\begin{Pro}\label{thm3}
If we encode spinor $\ket{u}=(\alpha, \beta)^T$, $\alpha,\beta\in\mathbb{C}$, into quaternion number by following
\begin{equation*}
    q := \mathrm{Re}(\alpha) + \mathrm{Im}(\alpha)\hat{\mathbf{i}} + \mathrm{Re}(\beta)\hat{\mathbf{j}} + \mathrm{Im}(\beta)\hat{\mathbf{k}}.
\end{equation*}
Then,
\begin{equation*}
     \mathrm{Re}\Big(q^*\underset{\vec{k}}{\Big\rvert}q\underset{\vec{p} - \vec{k}}{\Big\rvert}\Big)=\mathrm{Re}\Big(\bra{u(\vec{k})}\ket{u(\vec{p}-\vec{k})}\Big),\quad\forall\vec{k}\in\mathrm{BZ},
\end{equation*}
\end{Pro}
\begin{proof}
According to Property~(\ref{thm1}), it's suffice to show real part.
By the assumption, given $\vec{k}\in\mathrm{BZ}$ and let $\alpha = a+bi,~\beta=c+di$ with $a,b,c,d\in\mathbb{R}$, we have
\begin{equation}\label{eq:app_qq}
\begin{aligned}[b]
    q^*\underset{\vec{k}}{\Big\rvert}q\underset{\vec{p}-\vec{k}}{\Big\rvert} &= (a,b,c,d)^*\underset{\vec{k}}{\Big\rvert}(a,b,c,d)\underset{\vec{p}-\vec{k}}{\Big\rvert}\\
    &= (a,-b,-c,-d)\underset{\vec{k}}{\Big\rvert}(a,b,c,d)\underset{\vec{p}-\vec{k}}{\Big\rvert}\\
    &=\Big(a\underset{\vec{k}}{\Big\rvert}a\underset{\vec{p}-\vec{k}}{\Big\rvert}-(-b,-c,-d)\underset{\vec{k}}{\Big\rvert}\cdot(b,c,d)\underset{\vec{p}-\vec{k}}{\Big\rvert},~\vec{V}\underset{\vec{k}}{\Big\rvert}\Big)\\
    &=\Big(a\underset{\vec{k}}{\Big\rvert}a\underset{\vec{p}-\vec{k}}{\Big\rvert}+b\underset{\vec{k}}{\Big\rvert}b\underset{\vec{p}-\vec{k}}{\Big\rvert}+c\underset{\vec{k}}{\Big\rvert}c\underset{\vec{p}-\vec{k}}{\Big\rvert}+d\underset{\vec{k}}{\Big\rvert}d\underset{\vec{p}-\vec{k}}{\Big\rvert},~\vec{V}\underset{\vec{k}}{\Big\rvert}\Big)
\end{aligned}
\end{equation}
where $\vec{V}$ is the vector part of the quaternion product.
Notice that the second line above is a quaternion product, but the third line above is dot product between two vectors.
On the other hand
\begin{equation}\label{eq:app_uu}
\begin{aligned}[b]
    \bra{u(\vec{k})}\ket{u(\vec{p}-\vec{k})}&= \mqty(a-bi & c-di)\underset{\vec{k}}{\Big\rvert}\mqty(a+bi \\ c+di)\underset{\vec{p}-\vec{k}}{\Big\rvert}\\
    &=a\underset{\vec{k}}{\Big\rvert}a\underset{\vec{p}-\vec{k}}{\Big\rvert}+b\underset{\vec{k}}{\Big\rvert}b\underset{\vec{p}-\vec{k}}{\Big\rvert}+c\underset{\vec{k}}{\Big\rvert}c\underset{\vec{p}-\vec{k}}{\Big\rvert}+d\underset{\vec{k}}{\Big\rvert}d\underset{\vec{p}-\vec{k}}{\Big\rvert}+V
\end{aligned}
\end{equation}
where $V\in\mathbb{C}$ is the imaginary part.
It's obvious that Eq.~(\ref{eq:app_qq}) and Eq.~(\ref{eq:app_uu}) have exactly the same value in real part.
\end{proof}

\begin{Pro}
Given the quaternion number 
\begin{equation*}
    q := \mathrm{Re}(\alpha) + \mathrm{Im}(\alpha)\hat{\mathbf{i}} + \mathrm{Re}(\beta)\hat{\mathbf{j}} + \mathrm{Im}(\beta)\hat{\mathbf{k}},
\end{equation*}
from the spinor $\ket{u}=(\alpha, \beta)^T$, $\alpha,\beta\in\mathbb{C}$ under the gauge in Eqs.~(\ref{eq:gauge1}) and ~(\ref{eq:gauge2}).
If $h_3$ have the same sign in the entire BZ, then $F(\vec{p})= 0$. 
\end{Pro}
\begin{proof}
One can observe eigenvalues to conclude that 
\begin{equation}
\overline{h}=-\underline{h}=\sqrt{h_1^2+h_2^2+h_3^2},\quad\forall\vec{k}\in\mathrm{BZ}.
\end{equation}
From now on, we consider $h_3(\vec{k})>0$ for all $\vec{k}\in\mathrm{BZ}$, and therefore the spinor has the following form
\begin{equation*}
\begin{split}
    \ket{\overline{u}}
    &=\frac{1}{\sqrt{2\overline{h}(\overline{h}+h_3)}}\mqty(\overline{h}+h_3 \\ h_1+i h_2)\\
    &=\frac{1}{\sqrt{2\underline{h}(\underline{h}-h_3)}}\mqty(-\underline{h}+h_3 \\ h_1+ih_2)\\
    \ket{\underline{u}},
    &=\frac{1}{\sqrt{2\underline{h}(\underline{h}+h_3)}}\mqty(-h_1+i h_2 \\ \underline{h}+h_3).\\
\end{split}
\end{equation*}
After transforming the above two eigenstates into quaternions, we calculate the value of $\overline{q}^*\circledast\overline{q}-\underline{q}^*\circledast\underline{q}$ at $\vec{k}$:

\begin{widetext}
\begin{equation}\label{eq:app_f0}
\begin{aligned}[b]
F(\vec{p})& =\sum_{\vec{k}\in\mathrm{BZ}}
\Bigg(
\frac{1}{\sqrt{2\underline{h}(\underline{h}-h_3)}}(-\underline{h}+h_3,0, h_1, h_2)^*\underset{\vec{k}}{\bigg\rvert}
\frac{1}{\sqrt{2\underline{h}(\underline{h}-h_3)}}(-\underline{h}+h_3,0, h_1, h_2)\underset{\vec{p}-\vec{k}}{\bigg\rvert}\\
&\hspace{16em} - 
\frac{1}{\sqrt{2\underline{h}(\underline{h}+h_3)}}(-h_1,h_2, \underline{h}+h_3,0)^*\underset{\vec{k}}{\bigg\rvert}
\frac{1}{\sqrt{2\underline{h}(\underline{h}+h_3)}}(-h_1,h_2, \underline{h}+h_3,0)\underset{\vec{p}-\vec{k}}{\bigg\rvert}
\Bigg)\\
& = \sum_{\vec{k}\in\mathrm{BZ}}
\Bigg(
\frac{-\underline{h}+h_3}{\sqrt{2\underline{h}(\underline{h}-h_3)}}\underset{\vec{k}}{\bigg\rvert}
\frac{-\underline{h}+h_3}{\sqrt{2\underline{h}(\underline{h}-h_3)}}\underset{\vec{p}-\vec{k}}{\bigg\rvert}
+
\frac{h_1}{\sqrt{2\underline{h}(\underline{h}-h_3)}}\underset{\vec{k}}{\bigg\rvert}
\frac{h_1}{\sqrt{2\underline{h}(\underline{h}-h_3)}}\underset{\vec{p}-\vec{k}}{\bigg\rvert}
+
\frac{h_2}{\sqrt{2\underline{h}(\underline{h}-h_3)}}\underset{\vec{k}}{\bigg\rvert}
\frac{h_2}{\sqrt{2\underline{h}(\underline{h}-h_3)}}\underset{\vec{p}-\vec{k}}{\bigg\rvert}\\
&\qquad -
\frac{-h_1}{\sqrt{2\underline{h}(\underline{h}+h_3)}}\underset{\vec{k}}{\bigg\rvert}
\frac{-h_1}{\sqrt{2\underline{h}(\underline{h}+h_3)}}\underset{\vec{p}-\vec{k}}{\bigg\rvert}
-
\frac{-h_2}{\sqrt{2\underline{h}(\underline{h}+h_3)}}\underset{\vec{k}}{\bigg\rvert}
\frac{-h_2}{\sqrt{2\underline{h}(\underline{h}+h_3)}}\underset{\vec{p}-\vec{k}}{\bigg\rvert}
-
\frac{\underline{h}+h_3}{\sqrt{2\underline{h}(\underline{h}+h_3)}}\underset{\vec{k}}{\bigg\rvert}
\frac{-\underline{h}-h_3}{\sqrt{2\underline{h}(\underline{h}+h_3)}}\underset{\vec{p}-\vec{k}}{\bigg\rvert}
,\vec{V}\underset{\vec{k}}{\bigg\rvert}\Bigg)\\
&=\frac{1}{2}\sum_{\vec{k}\in\mathrm{BZ}}
\Bigg(
\frac{h_1}{\sqrt{\underline{h}(\underline{h}-h_3)}}\underset{\vec{k}}{\bigg\rvert}
\frac{h_1}{\sqrt{\underline{h}(\underline{h}-h_3)}}\underset{\vec{p}-\vec{k}}{\bigg\rvert}
-
\frac{h_1}{\sqrt{\underline{h}(\underline{h}+h_3)}}\underset{\vec{k}}{\bigg\rvert}
\frac{h_1}{\sqrt{\underline{h}(\underline{h}+h_3)}}\underset{\vec{p}-\vec{k}}{\bigg\rvert}\\
&\hspace{16em} +
\frac{h_2}{\sqrt{\underline{h}(\underline{h}-h_3)}}\underset{\vec{k}}{\bigg\rvert}
\frac{h_2}{\sqrt{\underline{h}(\underline{h}-h_3)}}\underset{\vec{p}-\vec{k}}{\bigg\rvert}
-
\frac{h_2}{\sqrt{\underline{h}(\underline{h}+h_3)}}\underset{\vec{k}}{\bigg\rvert}
\frac{h_2}{\sqrt{\underline{h}(\underline{h}+h_3)}}\underset{\vec{p}-\vec{k}}{\bigg\rvert}
,2\vec{V}\underset{\vec{k}}{\bigg\rvert}\Bigg),
\end{aligned}
\end{equation}
where $\vec{V}$ is the vector part of the above quaternion product at fixed $\vec{k}$ point.
Recalling Property~(\ref{thm1}) that shows the vector part has no contribution for real-valued function $F(\vec{p})$, it is suffice to calculate the real part of the quaternion product at fixed $\vec{k}$ point.

From Eq.~(\ref{eq:app_symm}), there are two points $\vec{k}=(k_x,k_y)$, and $\vec{k}'=(\pi-k_x,\pi-k_y)$ such that
\begin{equation}\label{eq:app_criteria}
\frac{h_i}{\sqrt{\underline{h}(\underline{h}-h_3)}}\underset{\vec{k}}{\bigg\rvert}
\frac{h_i}{\sqrt{\underline{h}(\underline{h}-h_3)}}\underset{\vec{p}-\vec{k}}{\bigg\rvert}
=
\frac{h_i}{\sqrt{\underline{h}(\underline{h}+h_3)}}\underset{\vec{k}'}{\bigg\rvert}
\frac{h_i}{\sqrt{\underline{h}(\underline{h}+h_3)}}\underset{\vec{p}-\vec{k}'}{\bigg\rvert}
,\quad \text{for~} i=1,~2,\: \text{and~} \vec{k},\vec{k}'\in\mathrm{BZ}.
\end{equation}
\end{widetext}
Therefore, the terms at $\vec{k}$ and at $\vec{k}'$ in Eq.~(\ref{eq:app_f0}) will cancel with each other.
Note that values at $\Gamma$ and at $(\pi,\pi)$ are zero in Eq.~(\ref{eq:app_f0}) since $h_1=h_2=0$ at these points in our model Eq.~(\ref{eq:HighChern}).
Thus, $F(\vec{p})=0$ if $h_3(\vec{k})>0$ for all $\vec{k}\in$  BZ.

Similarly, we assume $h_3(\vec{k})<0$ for all $\vec{k}\in$ BZ.
After calculation, the value of $F(\vec{p})$ is the same as Eq.~(\ref{eq:app_f0}).
Therefore, we can conclude that if $h_3$ has the same sign in the entire BZ, then $F(\vec{p})=0$.
\end{proof}


\bibliography{main}

\end{document}